\documentclass[%
 aip,
 amsmath,amssymb,
 reprint,%
]{revtex4-1}

\usepackage{graphicx}
\usepackage{dcolumn}
\usepackage{bm}
\usepackage[version=4]{mhchem} 
\usepackage{chemformula} 
\usepackage{subfigure}
\usepackage{makecell}
\usepackage{threeparttable} 

\setchemformula{kroeger-vink=true}

\makeatletter
\def\@email#1#2{%
 \endgroup
 \patchcmd{\titleblock@produce}
  {\frontmatter@RRAPformat}
  {\frontmatter@RRAPformat{\produce@RRAP{*#1\href{mailto:#2}{#2}}}\frontmatter@RRAPformat}
  {}{}
}%
\makeatother

\begin{document}

\preprint{AIP/123-QED}

\title{Defining ``Giant'' Electrostriction}
\author{Jiacheng YU, Pierre-Eymeric JANOLIN \\ Universit\'e Paris-Saclay, CNRS, CentraleSup\'elec, Laboratoire SPMS}

\date{\today}

\begin{abstract}
The recent discovery of ``giant'' electrostrictors has re-ignited the interest in electrostriction, an electromechanical coupling existing in all dielectrics but overshadowed by its linear counterpart: piezoelectricity. 
In this review, after a reminder of ``classical'' electrostriction, we propose a definition of ``giant'' electrostriction based on two empirical relations (``Newnam'' relation and one we posit). 
From this definition, we review previous reports on ``giant'' electrostrictors, to assess their nature. 
Focusing on the ones satisfying our definition, we compare their performances and characteristics. 
We also identify some of the hurdles to overcome before their adoption in the wide range of electromechanical applications, despite their fundamental and applicative interests.
\end{abstract}

\maketitle

\section{Introduction}
Electromechanical materials (piezoelectrics and electrostrictors) convert electric energy to mechanical energy (deformation or stress) and vice versa, thereby enabling sensors and actuators applications, for example in Micro-Electro-Mechanical Systems (MEMS)\cite{Bell2005,Liu2007} such as micro-machines, micro-switches, digital micro-mirrors, micro-valves and micro-pumps.
Inverse piezoelectricity is the most common electromechanical coupling for actuators, where a linear strain or stress is induced by an electric field or a polarization.
Another electromechanical coupling that can be used for actuators is electrostriction, albeit with a strain quadratic in electric field 
or polarization
. On the sensor side, direct piezoelectric effect is primarily used\cite{Anderson1994,Tonisch2006,Sharma2012}.
Electrostriction can also be used for sensing application, provided a bias electric field or pre-stress.
If piezoelectricity can only exist in dielectrics with a non-centrosymmetric structure, electrostriction occurs in all dielectrics without exception\cite{newnham2005properties}. 
The sign of the induced electrostrictive strain (positive, corresponding to stretching or expansion, or negative for compression) is independent of the field polarity. 
In most ceramics, electrostrictive stretching is observed along the direction of the external field\cite{uchino1980,zhao1995} whereas polymers generally exhibit the opposite effect, i.e. compression\cite{nakamura1971,Zhang1998}.

The electrostrictive effect is usually very small compared to commercial piezoelectrics. 
However, in 2012, Gd-doped ceria thin films\cite{Korobko2012} have been discovered to have an ``exceptionally large'' electrostriction coefficient, a behavior later also reported in ceramics\cite{Kabir2019}.
(Nb,Y)-stabilized bismuth oxide (\ch{Bi7Nb_{2-x}Y_xO_{2-2/x}})\cite{Yavo2016} and  \ch{La2Mo2O9} (LAMOX)\cite{Li2018} have been discovered to also exhibit ``giant'' electrostrictive performances. 
These ``giant'' electrostrictors generally have relative permittivity $\rm\varepsilon_r\approx30$, exhibit ``giant'' responses at low frequency (around 10\,Hz), and contain oxygen vacancies (\ch{V_{O}}) as candidate materials for  electrolytes in solid oxide fuel cells.

Lead halide perovskite single crystals methylammonium lead triiodide (\ch{MAPbI_3}) have also demonstrated ``giant'' (negative) electrostrictive responses 
at 100\,Hz\cite{chen2018}. 

In addition, soft nano-composites based on a polymer matrix and  carbon nanotubes (CNT) and liquid-crystalline graphene (rGO) are another type of remarkable electrostrictors as they have been reported to have an ``ultra-large'' response for frequencies between 0.1 and 100\,Hz\cite{luna2017,Yuan2018}.

``Giant'' electrostrictors offer a promising orientation toward the development of electromechanical applications not only due to their superior electromechanical properties, but also because they could provide a replacement for lead-based PZT that is at the heart of the vast majority of piezoelectric applications.

The issue is that not all these ``giant'' electrostrictors exhibit performances that are actually among the largest ones ever reported (this issue is not specific to recent reports as we shall show). In addition, electrostriction offers various metrics to gauge performance, thereby multiplying the opportunities for a material to stand out. Furthermore, as all dielectrics exhibit electrostriction, the pertinence of comparing e.g. inorganic single crystals and polymers is not a given. We shall nevertheless propose a definition of ``giant'' electrostriction suitable for all dielectrics, keeping the term ``giant'' even though the definition being based on what a ``normal'' electrostrictor is, it might have been more proper to call them ``anomalous''.

Hereafter, the first part is devoted to classical electrostriction: the definitions, expressions, and relationships between the various electrostrictive coefficients as well as the universal empirical relation that characterises the polarization electrostrictive coefficient $Q$. 
We shall demonstrate an equivalent relation for the field electrostrictive coefficient $M$, which is of greatest interest for applications. 
From these two relations, a quantitative criterion to qualify an electrostrictive response as ``giant'' shall be proposed. 

The second part of the paper reviews extensively the ``giant''  electrostrictors reported so far, their performances, and whether they satisfy the criterion we have defined to qualify as ``giant''. 

The third part assesses their interest both on the fundamental level and for applications; it also underlines the major hurdles to overcome to fully benefit from this effect. 
\subsection{Definition of electrostriction}

For linear electric and elastic materials, the polarization $P$ induced by an electric field $E$ is $P_i = \epsilon_0\chi_{ij} E_j$ (with $\epsilon_0$ the permittivity of vacuum and $\chi_{ij}$ the electric susceptibility) and Hooke's law gives the strain $x$ induced by a stress $X$: $x_{ij}=s_{ijkl} X_{kl}$ with $s$ the elastic compliance tensor. 
Electromechanical properties relate strain/stress to electric field/polarization or vice-versa. There are therefore four electromechanical coefficients (tensors). 
Similarly to the four piezoelectric tensors ($d$, $g$, $e$, and $h$), there are four electrostriction tensors. 

The first two electrostrictive tensors relate the strain generated by an electric field or a polarization. They
can be expressed\cite{Eury1999,Sundar1992,R.E.Newnham} as: 
\begin{equation}
\begin{split}
x_{ij}&=M_{ijkl} E_k E_l\\
x_{ij}&=Q_{ijkl} P_k P_l
\label{eq:def1}
\end{split}
\end{equation}
where 
$M_{ijkl}(\rm m^2/V^2)$ and $Q_{ijkl}(\rm m^4/C^2)$ are respectively the field- and polarization-electrostriction tensors.

In addition to $M$ and $Q$, the second pair of electrostrictive tensors ($m$ and $q$) are defined by the stress ($X_{ij}$) generated by an electric field or a polarization, respectively:
\begin{equation}
\begin{split}
X_{ij} &= m_{ijkl} E_k E_l\\
X_{ij} &= q_{ijkl} P_k P_l
\label{eq:def2}
\end{split}
\end{equation}

\begin{figure}[htbp]
    \centering
    \includegraphics[width=0.48\textwidth]{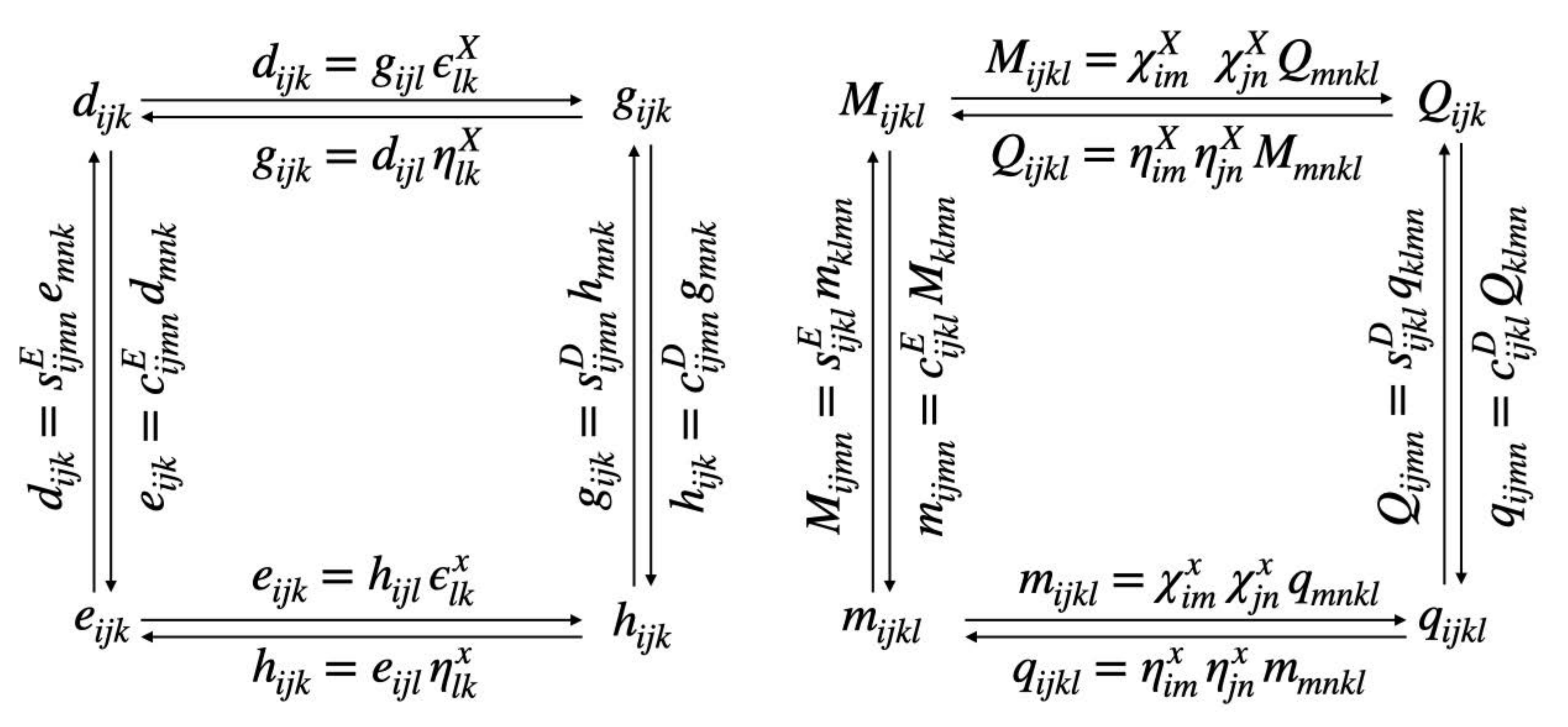}
    \caption{Relations among the four piezoelectric (left) and electrostrictive (right) tensors, via the elastic constant ($c_{ijkl}$) and compliance ($s_{ijkl}$) tensors, the electric susceptibility tensor ($\epsilon_0\chi_{ij}$), and its inverse ($\eta_{ij}$) in linear dielectric and elastic materials under isothermal or isentropic conditions. Exponents refer to the variables held constant.}
    \label{fig:Tensors}
\end{figure}

Relations between the four piezoelectric ($d$, $e$, $g$, and $h$), and the four electrostrictive tensors ($Q$, $q$, $M$, and $m$) are illustrated in Fig.\ref{fig:Tensors}. 

Practically speaking, the two components of the electric field ($E_{k}$ and $E_{l}$) are most conveniently applied along the same direction. 
This leads to the quadratic relations between strain or stress  and electric field or polarization. 
In addition, using Voigt notation, the expressions of the electrostrictive effects can be simplified:
\begin{equation}
\begin{split}
x_{i}&=M_{ik}E_k^2\\
x_{i}&=Q_{ik}P_k^2 \label{eq:square}
\end{split}
\end{equation}
with $i=1..6$ and $k=1..3$.

The difference between the effect described by the $M$ and $Q$ tensors appears under large fields.
If both $x$-$E$ and $x$-$P$ curves are quadratic under small amplitude of the electric field (i.e. are described by Eq.\eqref{eq:square}), the saturation of the polarization under large electric field in non-linear dielectrics leads to a saturation of the induced strain $x$. The $x$-$E$ curves is therefore not quadratic anymore, whereas the $x$-$P$ curve remains quadratic. Expanding the expression of the polarization as a function of the field to higher terms in Eq.\eqref{eq:square} leads to  Eq.\eqref{eq:nonlinear} where cubic, quartic etc. terms appear, whereas only the first (quadratic) term remains in linear dielectrics and leads to $M_{ijll}=Q_{ijkk}(\epsilon_0\chi_{kl})^2$.
\begin{equation}
\begin{split}
x_{ij}&=Q_{ijkk}P_k^2=Q_{ijkk}(\epsilon_0\chi_{kl}E_l+\epsilon_0\chi^{(2)}_{klm}E_lE_m+...)^2\\
&=Q_{ijkk}(\epsilon_0\chi_{kl})^2E_l^2+Q_{ijkk}(2\epsilon_0^2\chi_{kl}\chi^{(2)}_{klm})E_l^2E_m+...\\
&=M_{ijll}E_l^2+M^{(2)}E_l^2E_m+...\label{eq:nonlinear}
\end{split}
\end{equation}
with $\chi^{(2)}_{klm}$ the second-order electric susceptibility tensor. 

\begin{figure}
    \centering
    \label{fig:Saturation}
    \subfigure{\label{fig:Saturation(a)}
    \includegraphics[width=0.232\textwidth]{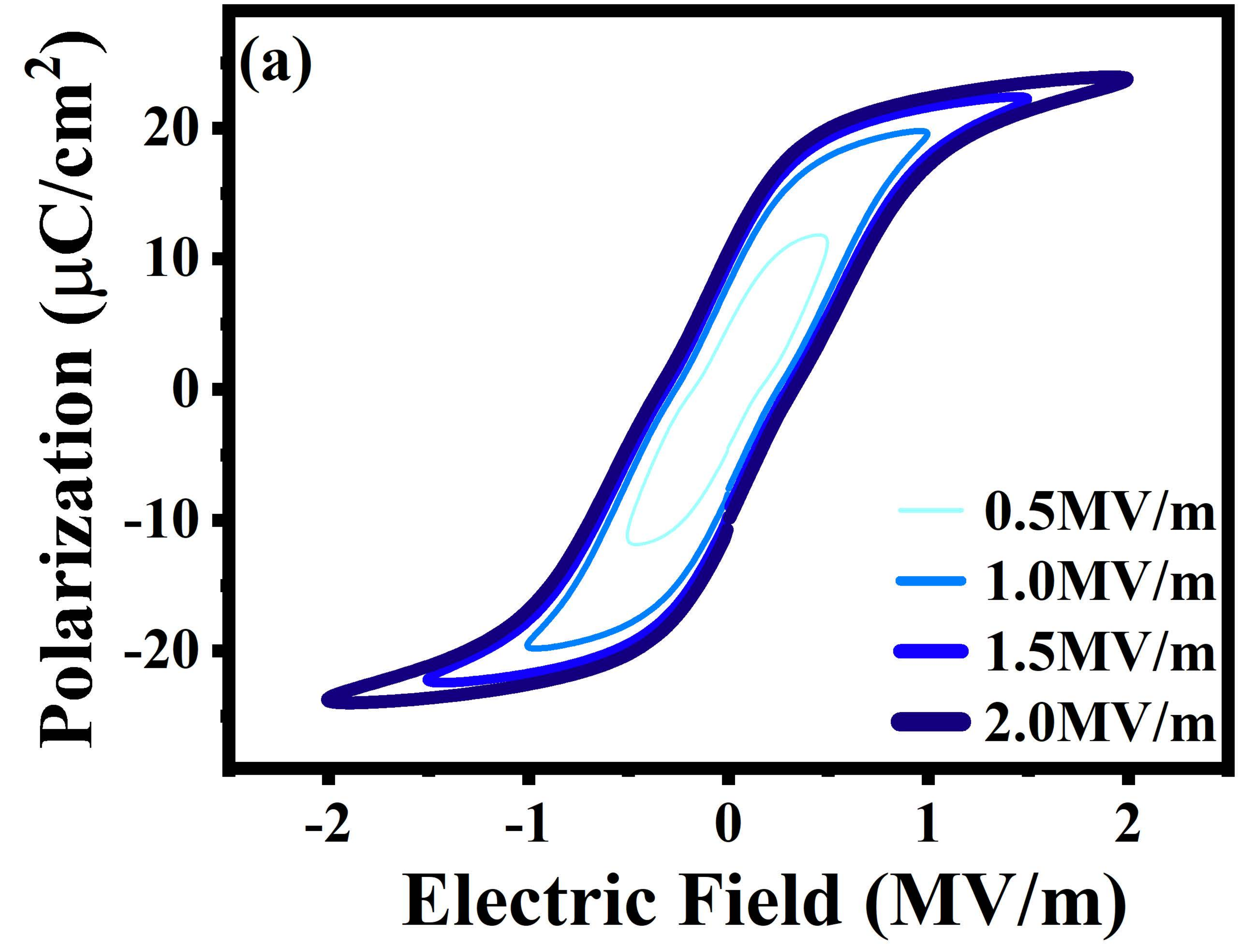}}
    \subfigure{\label{fig:Saturation(b)}
    \includegraphics[width=0.232\textwidth]{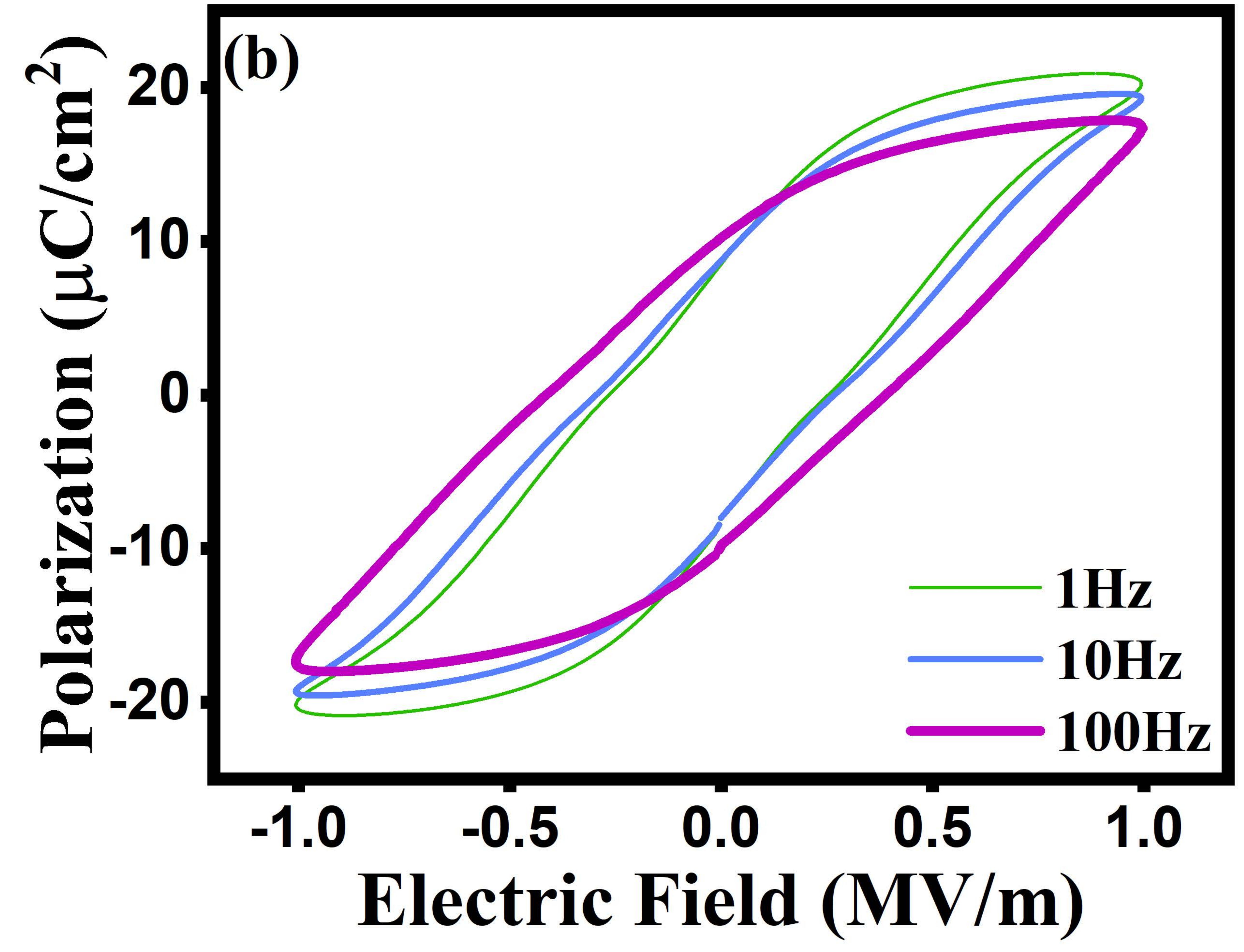}}
    \vfill
    \subfigure{\label{fig:Saturation(c)}
    \includegraphics[width=0.232\textwidth]{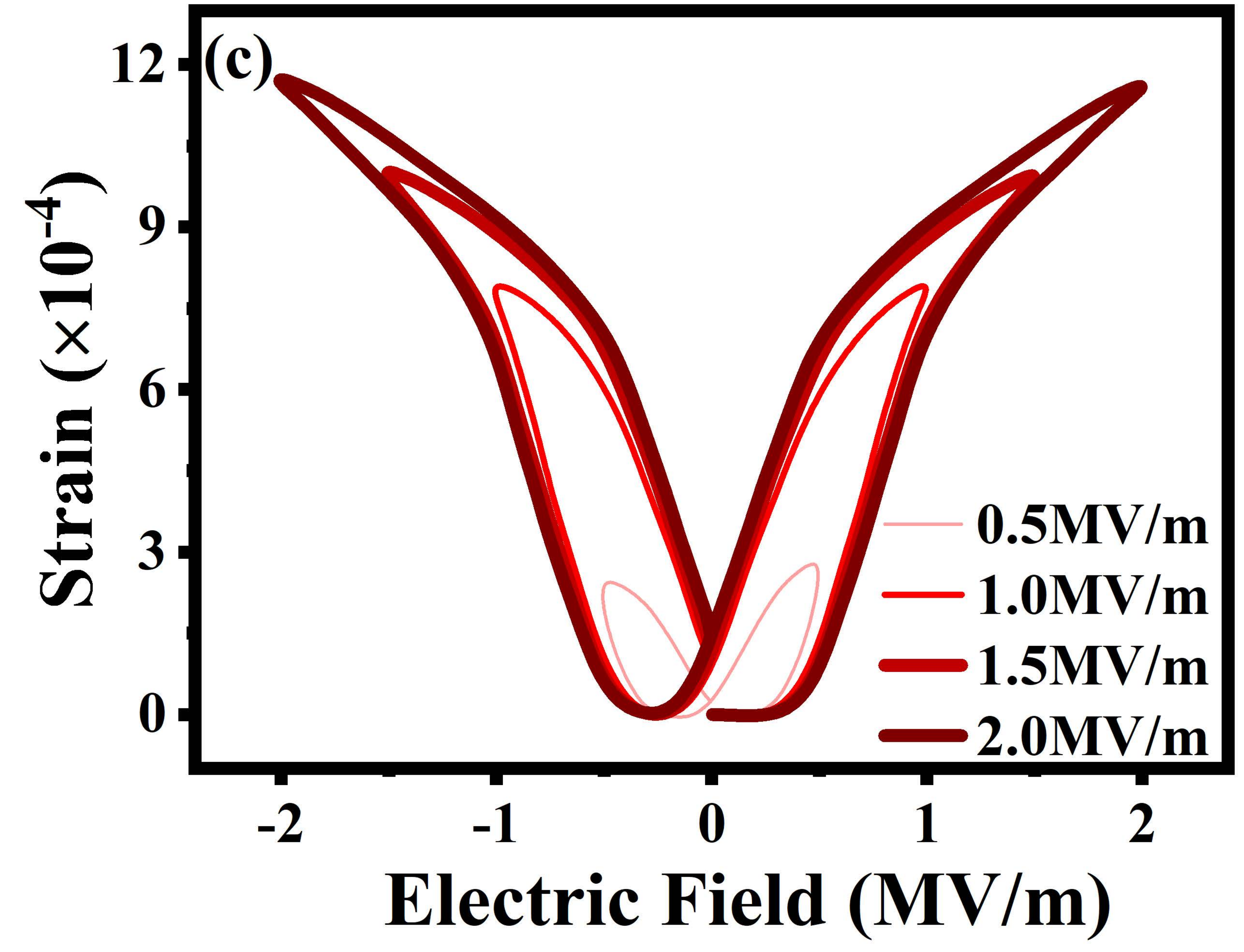}}
    \subfigure{\label{fig:Saturation(d)}
    \includegraphics[width=0.232\textwidth]{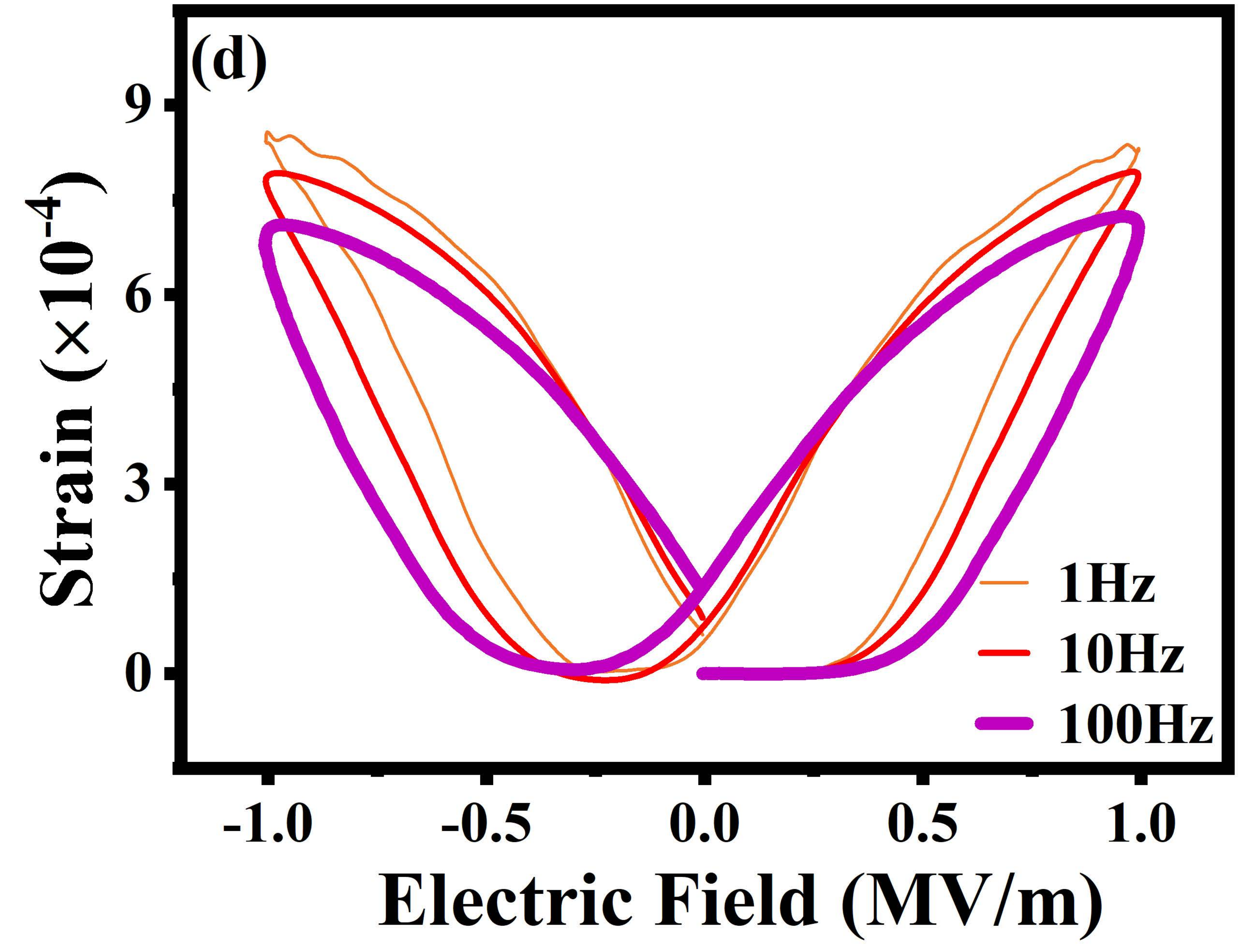}}
    \vfill
    \subfigure{\label{fig:Saturation(e)}
    \includegraphics[width=0.232\textwidth]{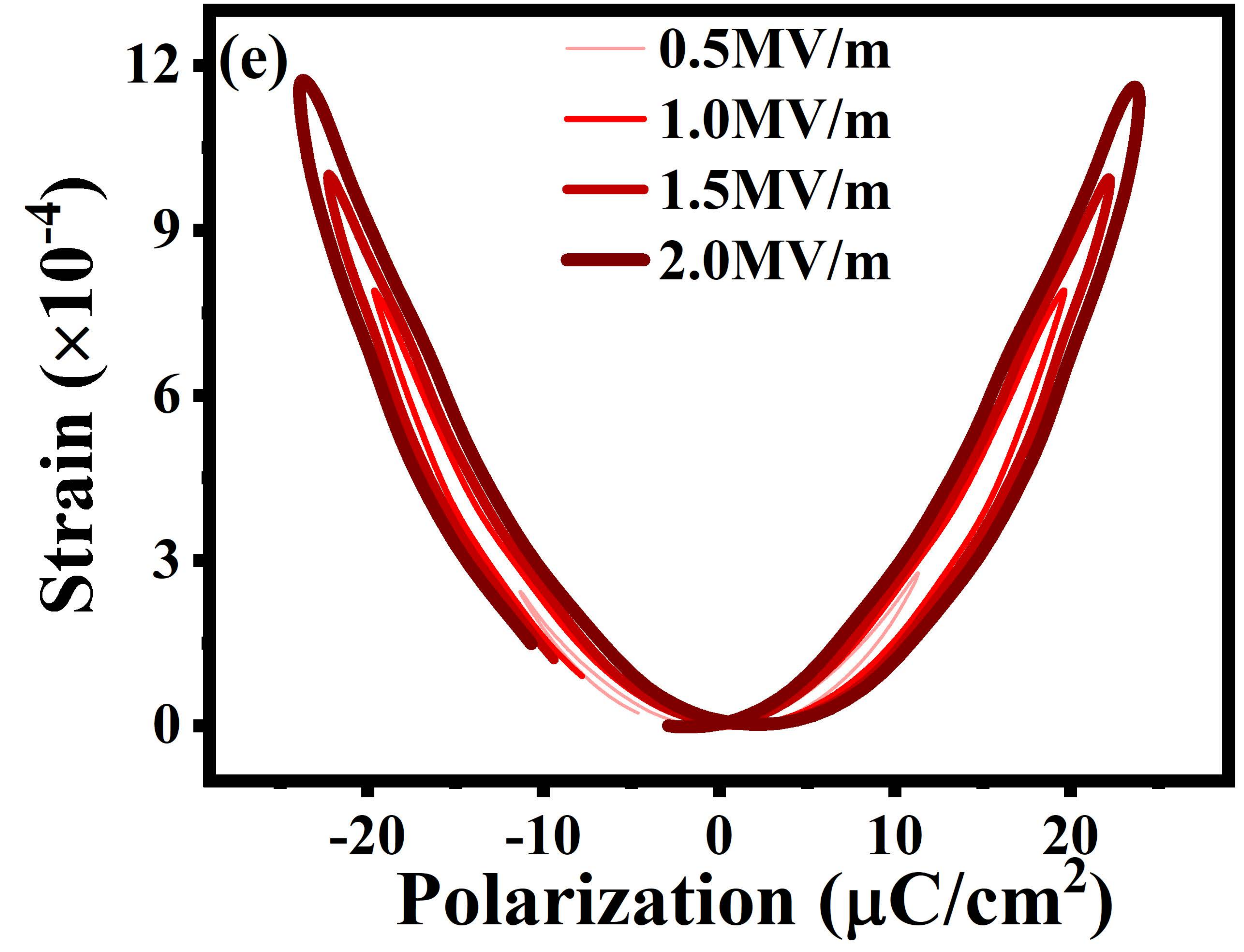}}
    \subfigure{\label{fig:Saturation(f)}
    \includegraphics[width=0.232\textwidth]{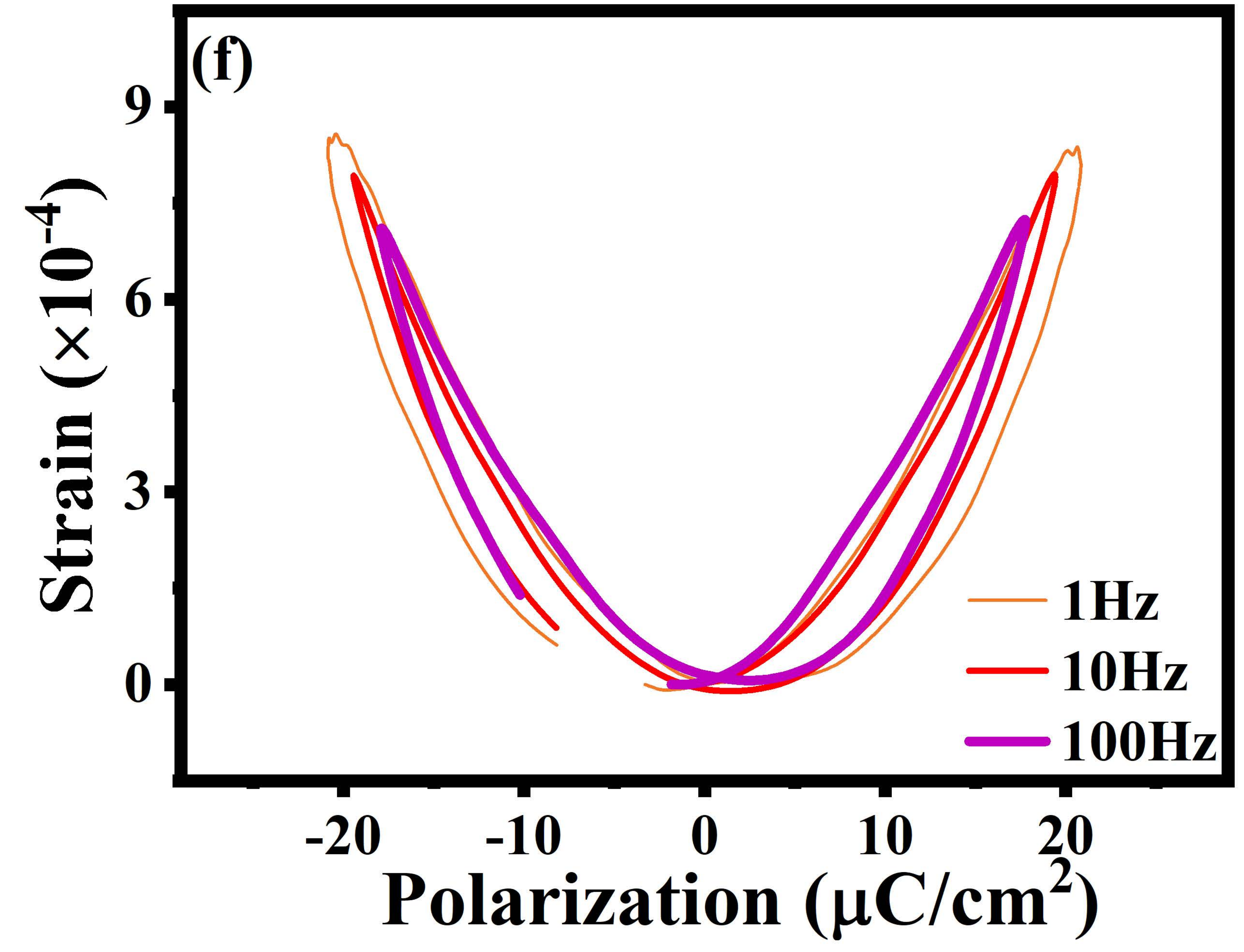}}
    \caption{The electric-field induced polarization for PMN-0.1PT crystals at 10\,Hz applied 0.5\,MV/m, 1\,MV/m, 1.5\,MV/m, 2\,MV/m respectively (a) and applied 1\,MV/m at 1\,Hz, 10\,Hz, 100\,Hz (b). The $x$-$E$ loops of PMN-0.1PT ceramic with respect to AC electric field at various amplitudes (c) and frequencies (d), which demonstrate the saturation with high electric field and larger electromechanical effect at lower frequencies. The $x$-$P$ loops PMN-0.1PT crystals at different AC electric field at 10\,Hz (e) and different frequencies at 1\,MV/m (f).}
\end{figure}

This behavior is illustrated in Figs.\ref{fig:Saturation(a)} and \ref{fig:Saturation(c)} on a ferroelectric relaxor PMN-0.1PT (\ch{(PbMg_{2/3}Nb_{1/3}O3)_{0.9}}-\ch{(PbTiO3)_{0.1})}\cite{kighelman2001,vikhnin2003,uchino1980,zhao1995} ceramic. 
The polarization-hysteresis loop (Fig.\ref{fig:Saturation(a)}) shows a saturation above $\approx$1\,MV/m. 
As a consequence the induced strain is only quadratic for low electric field ($x$-$E$ curve, Fig.\ref{fig:Saturation(c)}), whereas the strain remains quadratic in polarization ($x$-$P$ curve, Fig.\ref{fig:Saturation(e)}).
In addition, the phase difference between the polarization and the electric field is more pronounced in the $x$-$E$ curves than in the $x$-$P$ curve.
The evolution with frequency is also different. 
Figure \ref{fig:Saturation(d)} shows that the $M$ coefficient decreases with increasing frequency (the maximum strain decreases whereas the maximum field is constant).
On the contrary, the $Q$ coefficient increases (a given strain occurs for lower polarization values as the frequency increases), in agreement with the $P$-$E$ loop shown in Fig.\ref{fig:Saturation(b)}.
%
Even though $x$-$E$ loops (Fig.\ref{fig:Saturation(c)} and \ref{fig:Saturation(d)}) are not archetypal parabola, fast Fourier transform enables to evaluate the $M$ coefficient 
as electrostriction occurs at twice the frequency of the electric field.

\begin{figure}
    \centering
    \subfigure{\label{fig:PzEs(a)}
    \includegraphics[width=0.232\textwidth]{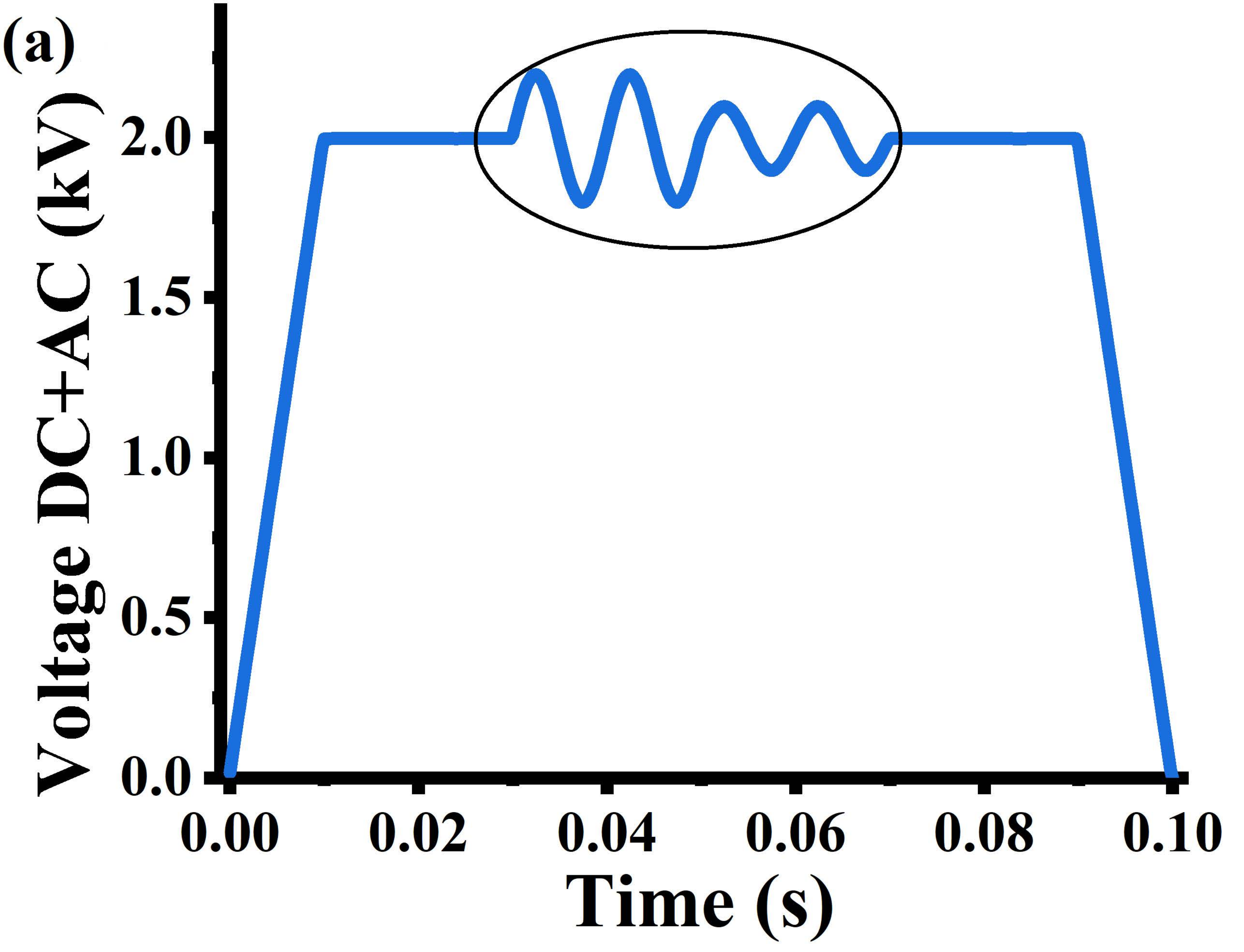}}
    \subfigure{\label{fig:PzEs(b)}
    \includegraphics[width=0.232\textwidth]{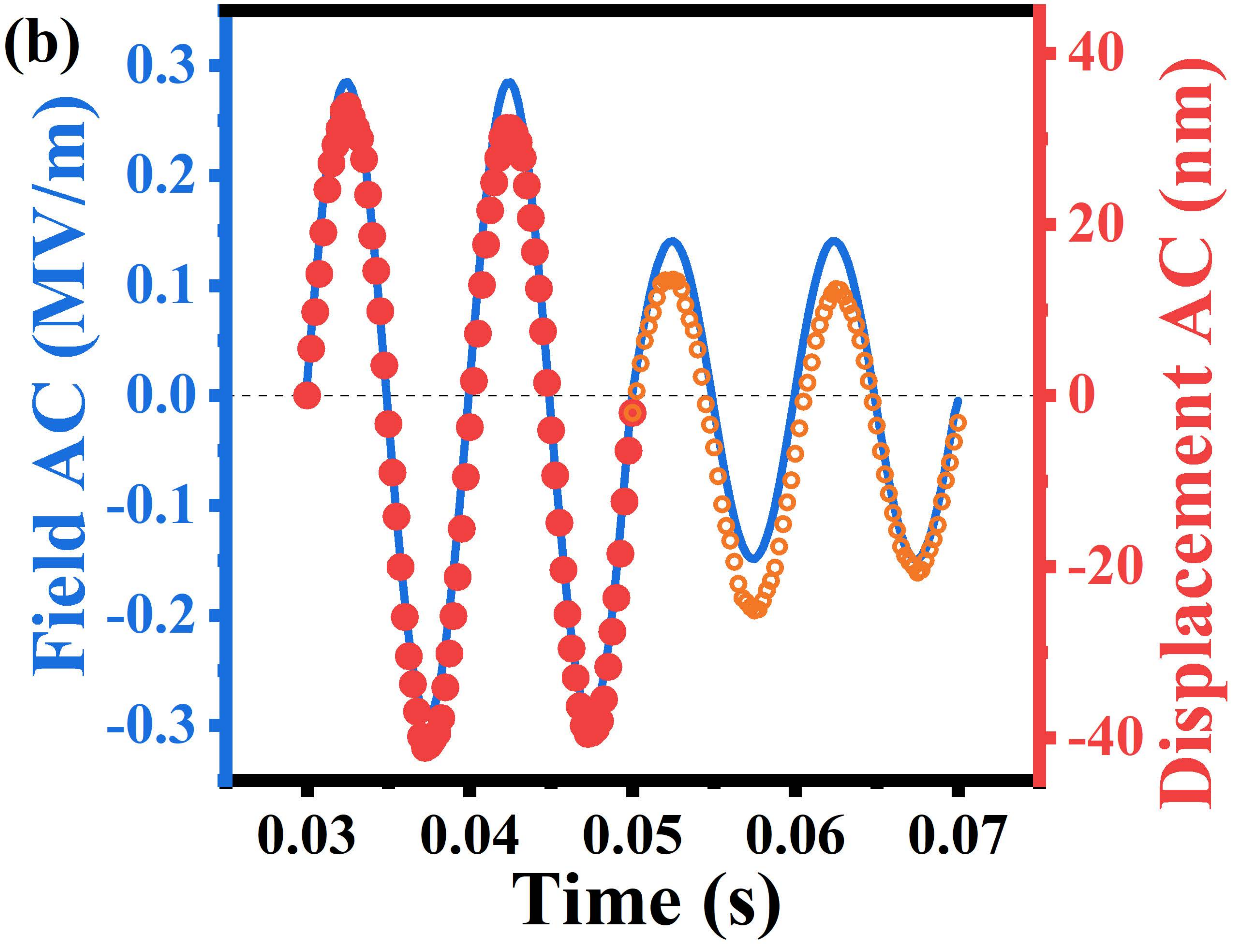}}
    \vfill
    \subfigure{\label{fig:PzEs(c)}
    \includegraphics[width=0.232\textwidth]{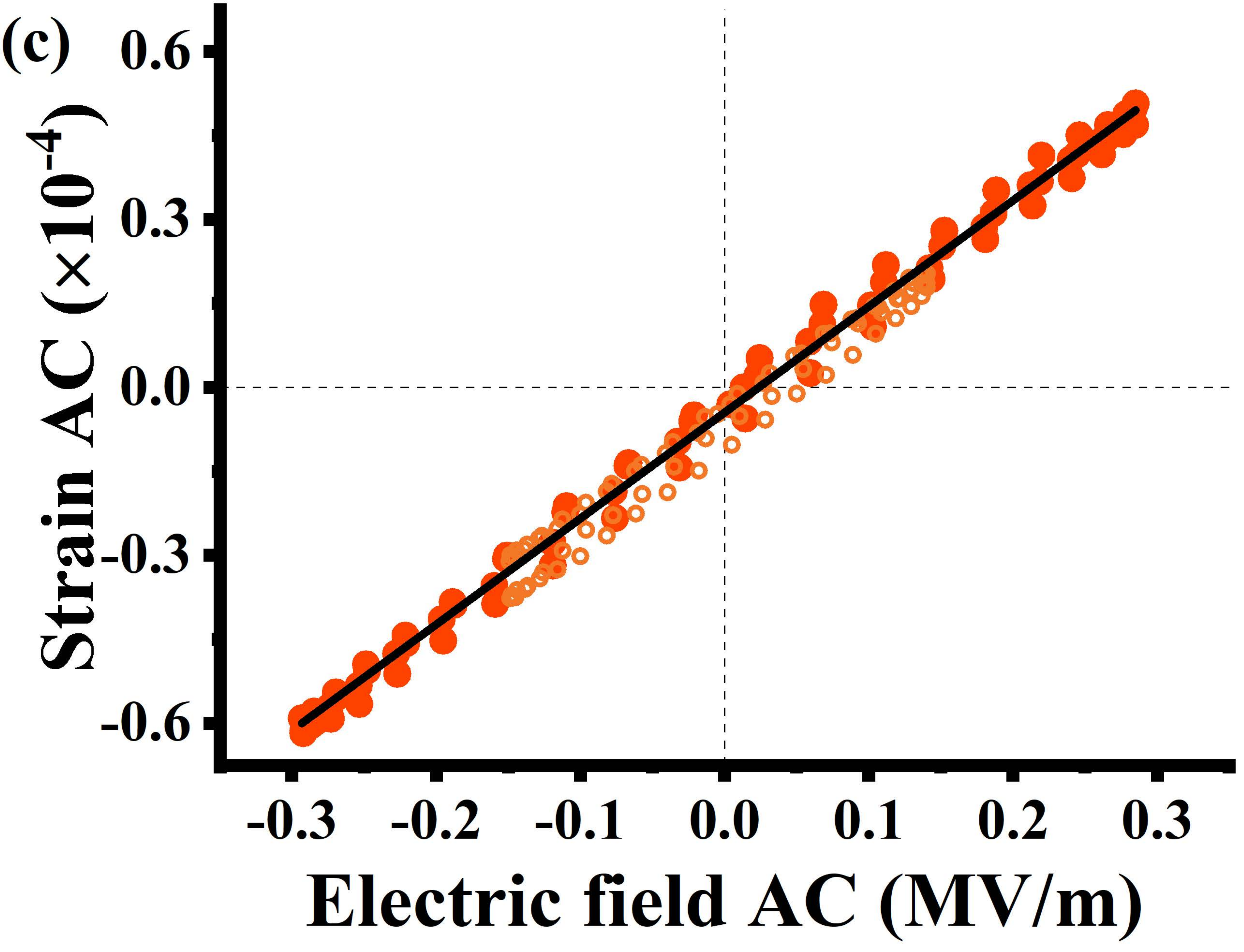}}
    \subfigure{\label{fig:PzEs(d)} 
    \includegraphics[width=0.232\textwidth]{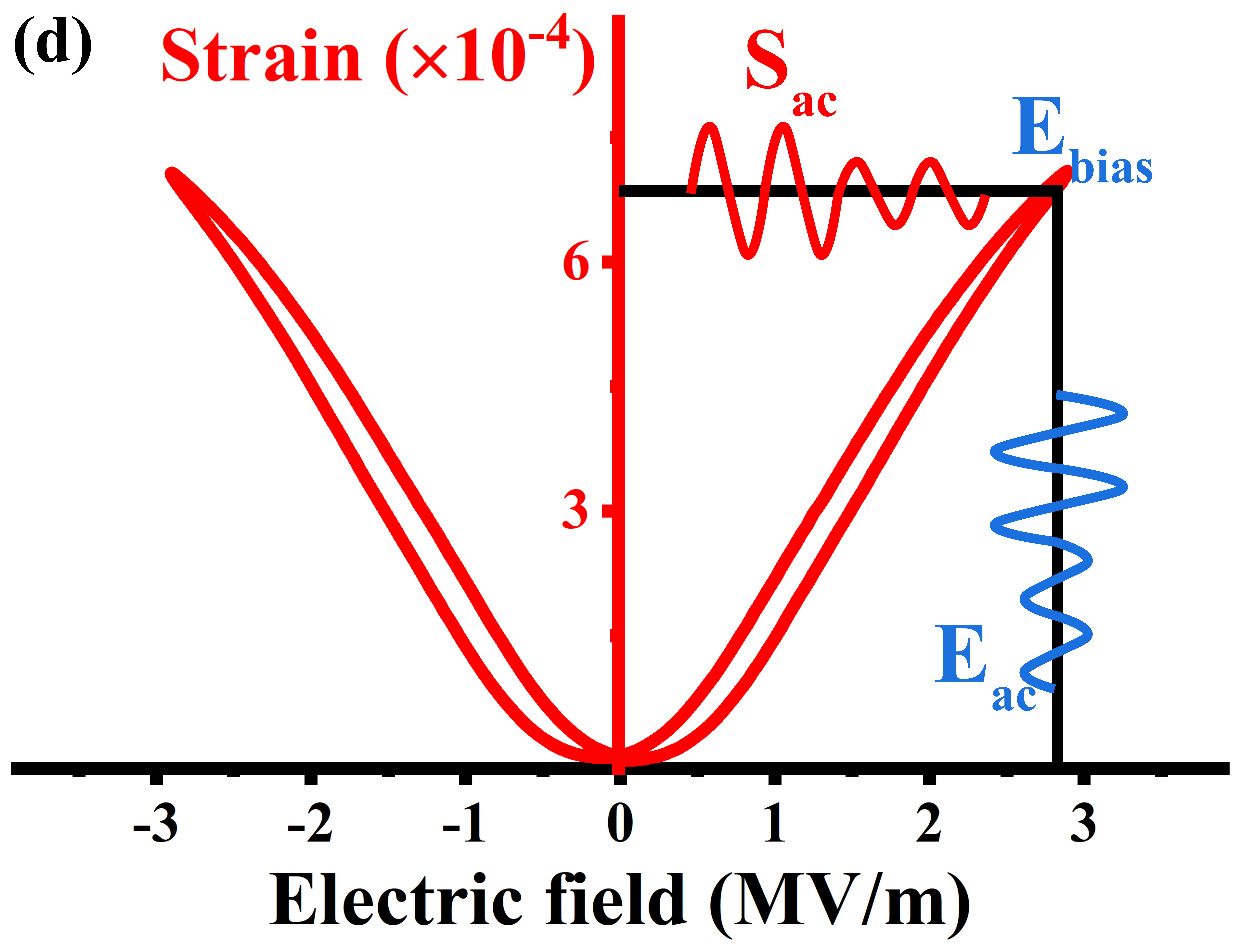}}
    \label{fig:PzEs}
    \caption{(a) Sinusoidal excitations (200Vpp and 100Vpp respectively) are added to a $dc$ voltage of 2\,kV on an electrostrictive (PMN-0.1PT crystal) sample. (b) The corresponding $ac$ electric field induces an $ac$ displacement at the same frequency and with a displacement amplitude proportional to the field, (c) corresponding to a linear ($ac$) strain response to an ($ac$) electric field; the slope provides the effective piezoelectric coefficient ($d_{33}^\mathrm{eff}$=182\,pm/V), independent of the amplitude of (small) $ac$ field. (d) Schematics of piezoelectricity originating under bias field\cite{li2014electrostrictive} from the measured electrostrictive response of a sample with  $M_{33}=(9.5\pm0.08)\times10^{-17}\rm\,m^2/V^2$.}
\end{figure}

Piezoelectricity and electrostriction are related.
For example, piezoelectricity stems from electrostriction in ferroelectric materials\cite{li2014electrostrictive,furukawa1990electrostriction}.
In addition, electrostrictors can exhibit an ``effective'' piezoelectric response when they are polarised by a $dc$ field\cite{Li2013}.
The effective piezoelectric coefficient is proportional to the product of the electrostrictive coefficient $M$ and bias field ($E_{bias}$ in Fig.\ref{fig:PzEs(d)})  
Fig.\ref{fig:PzEs(a)} shows the voltage profile with time that is applied on the sample, consisting of a $dc$ voltage (2\,kV) inducing a static strain. 
In order to measure the effective piezoelectric behavior, a sinusoidal voltage of $\pm$100\,V or $\pm$50\,V is applied at a frequency of 10\,Hz. 
The corresponding $ac$ field causes an oscillating displacement of the top surface of the sample (measured by laser interferometry) at the same frequency and with an amplitude proportional to the $ac$ field (see Fig.\ref{fig:PzEs(b)}). 
The proportionality and in-phase character result in a linear ($ac$) strain response to the applied $ac$ field (as illustrated in Fig.\ref{fig:PzEs(c)}), akin the one of a true piezoelectric.
The slope of this curve corresponds to the effective piezoelectric coefficient ($d^\mathrm{eff}$). 
%

As illustrated in Fig.\ref{fig:PzEs(d)}, the effective piezoelectric coefficient corresponds to the derivative of the $x$-$E$ curve around the strain induced by the \emph{dc} field, provided the \emph{ac} amplitude is small compared to the \emph{dc} one.
A large electrostriction coefficient $M$ therefore contributes to large effective piezoelectric coefficient, as shown in Eq.\eqref{eq:EPC} 
\begin{equation}
\begin{split}
d_{33}^\mathrm{eff}&=\frac{\partial x_{33}}{\partial E_{3}}=\frac{M_{33}\partial (E_{3(dc)}+E_{3(ac)})^2}{\partial E_{3(ac)}}\\
&=\frac{2M_{33}E_{3(dc)} \partial{ E_{3(ac)}}}{\partial E_{3(ac)}}=2M_{33}E_{3(dc)}
\label{eq:EPC}
\end{split}
\end{equation}
This is only valid for linear dielectrics. The saturation of the dielectric response induces a saturation of the strain response (see Fig.\ref{fig:PzEs(d)}). The corresponding effective piezoelectric coefficient (182\,pm/V) is therefore smaller than the value given by Eq.\eqref{eq:EPC} (549\,pm/V).

\subsection{The origin of electrostriction}

The origin of electrostriction is related to anharmonic effects. 
In other words, electrostriction coefficients are third derivatives (Eq.\ref{eq:MQmq}) of the free energy. 
Using the elastic Gibbs function $G_2$ and its differential form with respect to the free energy of vacuum:
\begin{equation}
\begin{split}
  G_2&=U-TS-E_iP_i\\
  dG_2&=-SdT+X_{ij}dx_{ij}-P_idE_i \label{eq:dG2}
\end{split}
\end{equation}
where $U$ the internal energy density ($J\cdot{m^{-3}}$), $T$ the temperature ($K$), $S$ the volume density of entropy ($J\cdot K\cdot m^{-3}$). 
The total differential related to the electric field 
is derived from Eq.\eqref{eq:dG2}:
\begin{equation}
\begin{split}
    \frac{d G_2}{d E_j}&=\frac{\partial}{\partial E_j}\left(\frac{dG_2}{dT}\right)_{x,E}dT+\frac{\partial}{\partial E_j}\left(\frac{dG_2}{dx_{kl}}\right)_{T,E}dx_{kl}\\
    &+\frac{\partial}{\partial E_j}\left(\frac{dG_2}{dE_i}\right)_{T,x}dE_i\\
    &=-\left.\frac{\partial S}{\partial E_j}\right|_{x,T}dT+\left.\frac{\partial X_{kl}}{\partial E_j}\right|_{T,x}dx_{kl}-\left.\frac{\partial P_i}{\partial E_j}\right|_{T,x}dE_i
\label{eq:dG2/dX}
\end{split}
\end{equation}
where the first term in Eq.\eqref{eq:dG2/dX} is the electrocaloric effect (inverse effect of pyroelectricity),  
the second term is the electromechanical coupling, and
the third term is purely dielectric. 
Combining Eqs.\eqref{eq:def2}, \eqref{eq:dG2} and \eqref{eq:dG2/dX}, the $m$ electrostrictive tensor can be related to the rate of change of the electric susceptibility with respect to strain in a linear dielectric:
\begin{equation}
\begin{split}
m_{ijkl}&=\frac{1}{2}\frac{\partial^3 G_2}{\partial E_i \partial E_j \partial x_{kl}}
=-\frac{1}{2}\frac{\partial^2 P_i}{\partial E_j \partial x_{kl}}\\&=-\frac{1}{2}\frac{\partial (\epsilon_0\chi_{ij})}{\partial x_{kl}}\\
\label{eq:MQmq}
\end{split}
\end{equation}
The three other electrostrictive coefficients can be derived from the electric Gibbs free energy $G_1$ ($Q_{ijkl}$), Gibbs free energy $G$ ($M_{ijkl}$) and Helmholtz free energy $A$ ($q_{ijkl}$). For linear dielectrics:
\begin{equation}
    \begin{split}
        Q_{ijkl}&=-\frac{1}{2}\frac{\partial^3 G_1}{\partial P_i \partial P_j \partial X_{kl}}=-\frac{1}{2}\frac{\partial (\epsilon_0\chi_{ij})^{-1}}{\partial X_{kl}}\\
        M_{ijkl}&=-\frac{1}{2}\frac{\partial^3 G}{\partial E_i \partial E_j \partial X_{kl}}=\frac{1}{2}\frac{\partial (\epsilon_0\chi_{ij})}{\partial X_{kl}}\\
        q_{ijkl}&=\frac{1}{2}\frac{\partial^3 A}{\partial P_i \partial P_j \partial x_{kl}}=\frac{1}{2}\frac{\partial (\epsilon_0\chi_{ij})^{-1}}{\partial x_{kl}}
    \end{split}
    \label{eq:MQmq2}
\end{equation}
Eq.\eqref{eq:MQmq} and \eqref{eq:MQmq2} show that electrostriction coefficients are related to the electric susceptibility (or its inverse) and strain or stress, which enables efficient \textit{ab initio} calculations\cite{tanner2020}. 
In addition, Eqs.\eqref{eq:MQmq} and \eqref{eq:MQmq2} can be rewritten as Eq.\eqref{eq:inverse}. This defines the inverse electrostrictive effect, that is the polarization or electric field induced by deforming a sample polarized by an electric field. 
An equivalent definition is the polarization (or electric field) induced by applying an electric field on (or polarising) a strained material.
\begin{equation}
  \begin{split}
\delta P_k=-2m_{ijkl}.\delta E_l.\delta x_{ij} \\
\delta E_k=-2q_{ijkl}.\delta P_l.\delta x_{ij}
\label{eq:inverse}
  \end{split}
\end{equation}

Direct electrostriction (Eqs.\eqref{eq:MQmq} and \eqref{eq:MQmq2}) enable applications as actuators, akin to inverse piezoelectricity. 
The existence of an inverse electrostriction (e.g. Eq.\eqref{eq:inverse}) is often overlooked, probably as it requires the application of a bias (electric or mechanical) field.
Similar to direct piezoelectricity, inverse electrostriction enables sensing. 


\subsection{General trend of hydrostatic electrostrictive coefficients}

To enable the comparison of electrostrictive properties between materials of different symmetry, an hydrostatic electrostriction coefficient $Q_h$ has been introduced\cite{R.E.Newnham,newnham2005properties,Uchino1981}.
$Q_h$ is defined as the quadratic relationship between the change of volume ($\Delta V/V$) and spontaneous polarization $P$ ($P^2=P^2_1+P^2_2+P^2_3$). From Eq.\eqref{eq:MQmq2}, $Q_h$ can be expressed as:
\begin{equation}
Q_h=\frac{\Delta V/V}{P^2}=-\frac{1}{2}\frac{\partial (\epsilon_0\chi)^{-1}}{\partial p}
\label{eq:Qh_d}
\end{equation}
where $p$ is hydrostatic pressure.

$Q_h$ characterizes the electrostrictive effects of isotropic or cubic materials, for instance $Q_h$=$Q_{iiii}$+$2Q_{iijj}$ for cubic materials, where $Q_{iiii}$ and $Q_{iijj}$ are respectively the longitudinal and transverse electrostriction coefficients.  
Hydrostatic electrostriction coefficients can be calculated in a cubic or isotropic sample from the measurement of longitudinal electrostriction alone using Eq.\eqref{eq:Qh}, provided that the Poisson's ratio ($\nu$) is known. 
\begin{equation}
 Q_{iijj}=-\nu Q_{iiii} \Rightarrow Q_h=(1-2\nu)Q_{iiii}
\label{eq:Qh}
\end{equation}

$Q_h$ is a function of the electric susceptibility, but not all materials exhibit the same dependence : perovskite oxides with comparable elastic compliances follow $Q_h \propto \chi$, linear dielectrics with a large variations of elastic compliance or low $\chi$ do not follow such proportionality, whereas soft organic dielectrics with a large elastic compliances exhibit in general larger $Q_h$\cite{R.E.Newnham}. 

\begin{figure}[htbp]
    \centering
    \includegraphics[width=0.47\textwidth]{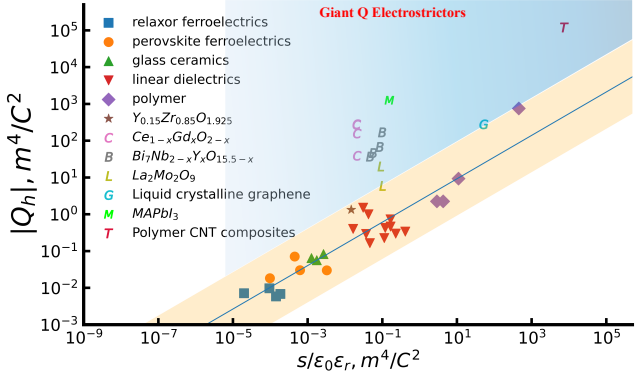}
    \caption{Empirical relation of hydrostatic electrostriction coefficient amplitude ($|Q_h|$), elastic compliance $s$,  and dielectric permittivity ($\epsilon_0\epsilon_r$). The data for classical electrostrictors are from Ref.\onlinecite{R.E.Newnham}, the blue lines corresponds to Eq.\eqref{eq:trend} and the yellow area corresponds to values within one order of magnitude of the line. Data point for Gd-doped ceria is from Ref.\onlinecite{Korobko2015}, bismuth oxide from Ref.\onlinecite{Yavo2016}, LAMOX from Ref.\onlinecite{Li2018}, liquid crystalline graphene from Ref.\onlinecite{Yuan2018},\ch{MAPbI3} from Ref.\onlinecite{chen2018}, Polymer CNT composites from Ref.\onlinecite{luna2017}}
    \label{fig:Empirical_EC}
\end{figure}

To reflect the electro-mechanical nature of electrostriction, a proxy combining the elastic and dielectric properties has been devised to estimate the amplitude of the hydrostatic electrostrictive response \cite{Uchino1981}. 
A ``universal'' linear relationship\cite{R.E.Newnham} was found between $\log(|Q_h|)$ and $\log(|s/ \varepsilon|)$ with $s$ the elastic compliance and $\epsilon$ the permittivity. We therefore propose to define ``classical'' electrostrictors as materials consistent with Equation \eqref{eq:trend}, illustrated in Fig. \ref{fig:Empirical_EC}. 
\begin{equation}
    |Q_h|\approx2.37\left(\frac{s}{\varepsilon_0\varepsilon_r}\right)^{0.59}   
\label{eq:trend}
\end{equation}

The empirical scaling law (Eq.\eqref{eq:trend}) qualitatively reflects the nature of most dielectrics. However, the units of $Q_h$ and $s/\varepsilon$ are both $\rm m^4/C^2$. It would therefore follow that the prefactor (2.37) is not dimensionless but has to be expressed in $\rm (m^4/C^2)^{-0.59}$. What is more this unit  probably depends on how many materials are used to fit Eq.\eqref{eq:trend}, as the slope of the fit would probably change. 
As a consequence, rather than carrying a physical meaning, Eq.\eqref{eq:trend} is rather useful as a proxy to evaluate qualitatively the electrostrictive properties of a given material, assuming it is a classical electrostrictor. 

The outliers in Fig.\ref{fig:Empirical_EC} (\emph{i.e.} materials out of the yellow area) exhibit electrostrictive properties well beyond the general trend defined by Eq.\eqref{eq:trend} (at least ten times larger). We therefore propose to only qualify as ``giant'', electrostrictors that exhibit an electrostrictive $Q_h$ coefficient more than one order of magnitude larger than the expected value from Eq.\eqref{eq:trend}. In Figure \ref{fig:Empirical_EC}, they lie above the yellow band. With this definition, ``giant'' electrostrictors are therefore \emph{anomalously} large i.e. ``giant'' compared to what one would expect based on their dielectric and elastic properties. This does not imply however that there are no ``classical'' electrostrictors with larger coefficients.

\begin{figure}[htbp]
    \centering
    \includegraphics[width=0.47\textwidth]{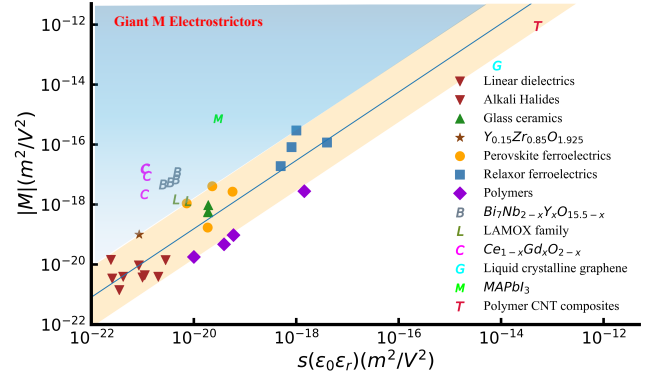}
    \caption{Empirical relation between the field electrostriction coefficient $M$ and the product of elastic compliance $s$ and dielectric permittivity $\epsilon_0\epsilon_r$. All data points are calculated from Fig.\ref{fig:Empirical_EC}.}
    \label{fig:Mse}
\end{figure}

Eqs.\eqref{eq:def1} or \eqref{eq:square} show that the electrostrictive  strain is related to the product of the electrostrictive coefficient $Q$ by the polarization squared. 
Therefore, if the permittivity (hence the susceptibility) is modest and/or saturates at low field, a large electric field will not result in a large polarization and the electrostrictive strain will remain of limited amplitude. 
The field response (hence the $M$ coefficient) depends on the ability of accumulating charges under electric field and of deforming, which correspond respectively to the  permittivity $\varepsilon$ and elasticity $s$. 
We therefore posit that the amplitude of the $M$ coefficient is correlated to the product of the elastic susceptibility $s$ by the permittivity $\varepsilon_0\varepsilon_r$, both quantities are in $\rm m^2/V^2$.
Fig.\ref{fig:Mse} demonstrates the general trend of the $M$ coefficient with respect to the product $s(\epsilon_0\epsilon_r)$ for the same set of materials as in Fig.\ref{fig:Empirical_EC}. Eq.\eqref{eq:Mtrend} estimates numerically this relation from a fit excluding ` ``giant'' $Q$ electrostrictors:
\begin{equation}
    |M|\approx10^{4}(s\varepsilon_0\varepsilon_r)^{1.14}   
\label{eq:Mtrend}
\end{equation}
As for Eq.\ref{eq:trend}, the power of $\rm s\epsilon_0\epsilon_r$ being different from 1, the pre-factor would have to be in a unit ensuring the homogeneity of the relation and would probably depend on the number of materials used in the fit. Eq.\eqref{eq:Mtrend} (as Eq.\eqref{eq:trend}) is therefore not a physical ``law'' but a proxy to qualitatively evaluate the electrostrictive properties of very different materials.  

Nevertheless, as Eq.\eqref{eq:trend} was used to put forward the definition of giant electrostrictors based on their $Q$ coefficient, we propose Eq.\eqref{eq:Mtrend} as the basis for definition of ``giant'' $M$ electrostrictors: materials lying above the yellow band in Fig.\ref{fig:Mse} are more than order of magnitude larger than the value estimated from Eq.\eqref{eq:Mtrend} and qualify as ``giant'' ($M$) electrostrictors.

%

Experimental data on strain, permittivity, and elasticity 
may not have been measured on the sample or under the same experimental conditions. As a consequence, there is a dispersion of the data points in Fig.\ref{fig:Empirical_EC} and Fig.\ref{fig:Mse}. 
In addition, the measured strain is always the result of the electrostrictive response and Maxwell stress (Eq.\ref{eq:Max}), the latter resulting from the electrostatic attraction between electrodes. 
\begin{equation}
x_M=-\frac{1}{2}s\varepsilon_0\varepsilon_r E^2
\label{eq:Max}
\end{equation}
Maxwell stress is always negative (compressive), quadratic in  electric field, and its coefficient is half of the product of permittivity by the elastic compliance. It may be not negligible, especially in the case of materials with large permittivities (e.g. ferroelectric relaxors) or elastic compliances (e.g. polymers). Whether Maxwell strain has been removed or not from the measured strain is not always explicitly mentioned in the reports, potentially adding to the dispersion of the experimental results. 

\begin{figure}[htbp]
    \centering
    \includegraphics[width=0.47\textwidth]{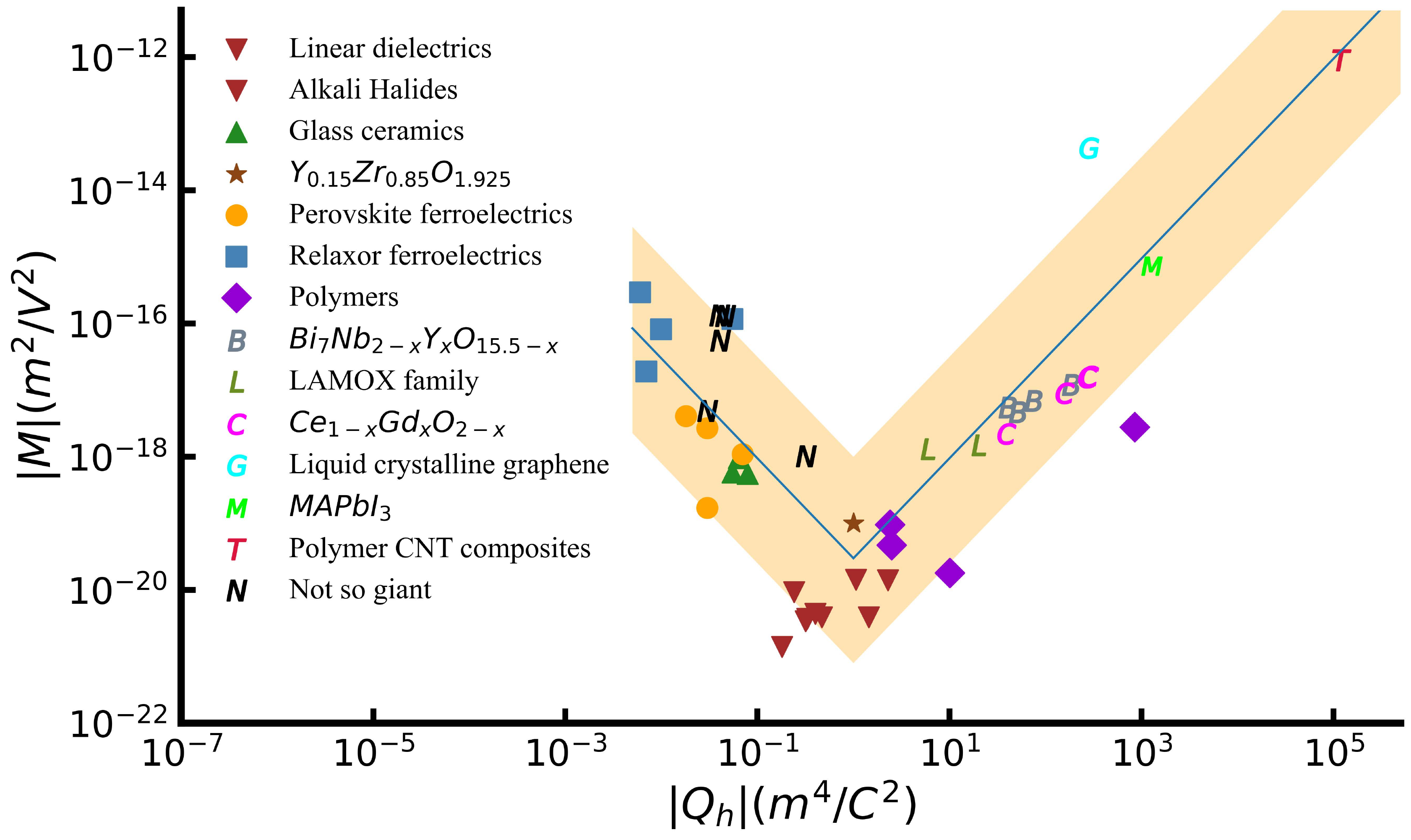}
    \caption{Hydrostatic polarization ($|Q_h|$) vs field ($|M|$) electrostriction coefficient showing that the two values are not correlated.}
    \label{fig:MQh}
\end{figure}

Figures \ref{fig:Empirical_EC} and \ref{fig:Mse} reveal that if most of the ``giant'' $Q$ electrostrictors also qualify as ``giant'' $M$ electrostrictors, there are exceptions (e.g. the two polymer-based composites). 
A comparison of the polarization ($|Q_h|$) and field ($|M|$) electrostriction coefficients amplitudes is shown in Fig.\ref{fig:MQh}. This relation is not monotonic but rather exhibits a ``V''-shape where all values fit into a band of $\pm$1 order of magnitude, including ``giant'' electrostrictors, with the only exception of liquid crystalline graphene nanocomposites.

\section{Reported ``Giant'' Electrostrictors}

Figures \ref{fig:Empirical_EC} and \ref{fig:Mse} also show that some ``classical'' electrostrictors exhibit electrostrictive coefficients that are larger than the ones of ``giant'' electrostrictors. 
For example, relaxor ferroelectrics (the workhorse of electrostrictive applications) exhibit $M$ coefficients comparable or larger than most ``giant'' electrostrictors. 
This is mainly due to their very large permittivity. 
Increasing the dielectric properties of electrostrictors is an  extensively investigated\cite{li2014electrostrictive} approach to improve $M$.
``Giant'' electrostrictors therefore exhibit an \textit{anomalously} large response (i.e. much larger than expected from Eqs.\eqref{eq:trend} and \eqref{eq:Mtrend}) rather than an \textit{absolutely} large response as some classical electrostrictors, especially polymers and composites, can exhibit larger values of electrostrictive coefficients.

Classical electrostrictors have been investigated in several reviews (see e.g. Refs.\onlinecite{uchino1986,bauer2012}). Here we shall therefore restrict our scope to the electrostrictors that have been reported as ``giant''. 

In 2012, Gd-doped ceria, an oxygen ionic conductor used as electrolyte in solid oxide fuel cells, was reported to exhibit a giant electrostrictive effect. 
It can generate stresses in excess of 500\,MPa\cite{Korobko2012}. 
The upper limit of modern inorganic piezoelectrics and electrostrictors does not exceed 150\,MPa (see Ref.\onlinecite{Korobko2012} and references therein).

Four years later, the hydrostatic polarization electrostriction coefficient $Q_h$ of (Y, Nb)-stabilized $\rm\delta$-$\rm Bi_2O_3$ was measured to be in the range of 40-182\,$\rm m^4/C^2$, well above the expected value\cite{Yavo2016} (see Fig.\ref{fig:Empirical_EC}).

Both materials crystallize in the fluorite structure and exhibit anelasticity. However, not all fluorites exhibit such giant electrostriction. Fluorite ionic conductor like Y-doped zirconia (\ch{Y_{0.15}Zr_{0.85}O_{1.925}}) exhibits a modest $ Q_h\rm\sim1 m^4/C^2$, hundred times smaller than $\rm\delta$-$\rm Bi_2O_3$\cite{Yavo2016}, despite the anelasticity reported in 8\% mol and 10\% mol Y-doped zirconia\cite{gao2014}.

In 2018, monoclinic \ce{La2Mo2O9} and cubic LAMOX (\ce{La2MoWO9})\cite{Li2018} were reported to exhibit $Q_h$ values of about 20 $\rm m^4/C^2$, showing that the fluorite structure was not a prerequisite, nor mandatory for ``giant'' electrostriction to occur. 

All these materials have in common a large concentration of oxygen vacancies distorting the crystal. It is however not enough to induce a large electrostrictive response. For example, \ch{Y_{0.15}Zr_{0.85}O_{1.925}} has an oxygen vacancies concentration of 3.75\% but does not exhibit a giant electrostrictive\cite{Yavo2018}.

The same year, the $M$ electrostrictive coefficient of liquid crystalline graphene nanocomposite was found\cite{Yuan2018} to be $M\sim$10$^{-14}\,$m$^2$/V$^2$ at 0.1\,Hz with a relative permittivity of $\varepsilon_r\sim10^3$. Despite being very large, such a value does not qualify these materials as ``giant'' electrostrictors. The same holds true for polymer carbon nanotube composites\cite{luna2017} that were reported to have an even larger  $M\sim$10$^{-13}\,$m$^2$/V$^2$ at 100\,Hz with a relative permittivity $\varepsilon_r=300$. This is the highest value reported so far. 
Polymer-matrix based nanocomposites may become a prolific area for the discovery of new electrostrictors.

Lead halide perovskites have demonstrated outstanding performance in photovoltaics. Still in 2018, a methylammonium lead triiodide (\ce{MAPbI3}) single crystal was reported to exhibit a compressive strain about 1\%, with $M\sim7.3\times$10$^{-16}\,$m$^2$/V$^2$ at 10\,Hz, qualifying it as a ``giant'' $M$ (and $Q$) electrostrictor.
Other lead halide perovskites (such as \ce{CsPbI3} and \ce{MAPbBr3} single crystals) electromechanical responses were also measured at $M\sim8\times$10$^{-17}\,$m$^2$/V$^2$ and $M\sim9\times$10$^{-16}\,$m$^2$/V$^2$ respectively\cite{chen2018}.
Table \ref{table:GE} summarizes all the ``giant'' electrostrictors reported until now, with their electrostrictive $M$ and $Q_h$ coefficients, their relative permittivity, the maximum field applied and the frequency of the maximum electrostrictive strain. 
Table \ref{table:GE} provides as well the amplitude of the effective (longitudinal) piezoelectric coefficient ($\rm |d^{eff}|$) and the figure of merit for energy harvesting applications. 
These values are either extracted from the publications or calculated from available experimental values. 
For comparison, selected ``normal'' electrostrictors and piezoelectrics are also present in Table \ref{table:GE}.

\begin{table*}
\begin{ruledtabular}
    \caption{Maximum responses of ``giant'' electrostrictors} 
    \label{table:GE}
    \begin{tabular}{ccccccccc}
    \textbf{\makecell{Giant\\electrostrictors}} & 
    \textbf{\makecell{Maximum field\\(MV/m)}} & 
    \textbf{\makecell{Frequency\\(Hz)}} & 
    \textbf{\makecell{$\rm|M|$\\$(10^{-18}\rm\,m^2/V^2)$}} & 
    \textbf{\makecell{$\rm|Q|$\\$(\rm m^4/C^2)$}} & 
    \textbf{$\varepsilon_r$} & 
    \textbf{\makecell{$\rm |d^{eff}|$ $^\#$\\$(\rm pm/V)$}} &
    \textbf{\makecell{$\rm d^{eff}\cdot g^{eff}$\\$(10^{-12}\rm\,m^2/N)$}} &
    \textbf{Ref}\\\hline
    \makecell{\ch{Ce_{0.8}Gd_{0.2}O_{1.9}}\\thin film} & 6\,$^a$& 0.01\,$^a$ & 9\,$^a$ & 157\,$^a$ & 28\,$^a$& 108\,$^b$ & 47.0 & \cite{Korobko2012}\\ 
    \makecell{\ch{Ce_{0.9}Gd_{0.1}O_{1.95}}\\thin film} & 6\,$^a$ & 0.01\,$^a$ & 16\,$^a$ & 282\,$^a$ & 28\,$^a$ & 192\,$^b$ & 148.7 & \cite{Korobko2015} \\
    \makecell{\ch{Ce_{0.9}Gd_{0.1}O_{1.95}} \\ceramics} & 0.374\,$^b$ & 0.15$\sim$9\,$^a$ & 21.7\,$^a$ & 3797\,$^b$ & 28\,$^a$ & 16.2\,$^b$ & 1.1 & \cite{Yavo2018}\\
    \makecell{\ch{Bi_7Nb_{0.4}Y_{1.6}O_{13.9}} \\ceramics} & 1.1\,$^a$ & 20$\sim$1200\,$^a$ & 12\,$^a$ & 273\,$^a$ & 24.6\,$^a$ & 26.4\,$^b$ & 3.2 & \cite{Yavo2016}\\
    \makecell{\ch{Bi_7Nb_{0.8}Y_{1.2}O_{14.3}} \\ceramics} & 1.1\,$^a$ & 20$\sim$1200\,$^a$ & 6.9\,$^a$ & 142\,$^a$ & 26\,$^a$ & 15.1\,$^b$ & 1.0 & \cite{Yavo2016}\\
    \makecell{\ch{Bi_7Nb_{1.2}Y_{0.8}O_{14.7}} \\ceramics} & 1.1\,$^a$ & 20$\sim$1200\,$^a$ & 5.5\,$^a$ & 83\,$^a$ & 30\,$^a$ & 12.1\,$^b$ & 0.55 & \cite{Yavo2016}\\
    \makecell{\ch{Bi_7Nb_{1.6}Y_{0.4}O_{15.1}} \\ceramics} & 1.1\,$^a$ & 20$\sim$1200\,$^a$ & 4.7\,$^a$ & 118\,$^a$ & 23.5\,$^a$ & 10.3\,$^b$ & 0.51 & \cite{Yavo2016}\\
    \makecell{\ch{La_{2}Mo_{2}O_{9}} \\ceramics} & 11\,$^b$ & 10\,$^a$ & 1.5\,$^a$ & 20\,$^b$ & 28 \,$^a$& 33\,$^b$ & 4.4& \cite{Li2018}\\
    \makecell{\ch{La_{2}MoWO_{9}} \\ceramics} &  11\,$^b$ & 10\,$^a$  & 1.3\,$^a$  & 5.9\,$^b$  & 48\,$^a$  & 28.6\,$^b$ & 1.9& \cite{Li2018}\\
    \makecell{\ch{MAPbI3}\\single crystals} & 3.7\,$^a$ & 100\,$^a$ & 730\,$^a$ & $\rm |Q_{33}|$=1266\,$^b$ & 50\,$^a$ & 5402\,$^b$ & 65917 & \cite{chen2018}\\
    \makecell{\ch{MAPbBr3}\\single crystals} & 2\,$^a$ & 10\,$^a$ & 900\,$^a$ & $\rm |Q_{33}|$=1417\,$^b$ & 90\,$^a$ & 3600\,$^b$ & 16264 & \cite{chen2018,gao2019}\\\hline\hline
    \textbf{Electrostrictors} & 
    \textbf{\makecell{Maximum field\\(MV/m)}} & 
    \textbf{\makecell{Frequency\\(Hz)}} & 
    \textbf{\makecell{$\rm|M|$\\$(10^{-18}\rm\,m^2/V^2)$}} & 
    \textbf{\makecell{$\rm|Q|$\\$(\rm m^4/C^2)$}} & 
    \textbf{$\varepsilon_r$} & 
    \textbf{\makecell{$\rm |d^{eff}|$ $^\#$\\$(\rm pm/V)$}} &
    \textbf{\makecell{$\rm d^{eff}\cdot g^{eff}$\\$(10^{-12}\rm\,m^2/N)$}} &
    \textbf{Ref}\\\hline
    \makecell{0.5\% CNT\\ polymer composites} & \makecell{Not mention} & 100\,$^a$ & $\rm 9\times10^{5}$\,$^a$ & $\rm |Q_{33}|$=$\rm 1.2\times10^{5}$\,$^a$\ & 300\,$^a$ &-- & -- & \cite{luna2017}\\
    \makecell{rGO/PMDS\\nanocomposites} & \makecell{Not mention} & 0.1\,$^a$ & $\rm 4\times10^{4}$\,$^a$ & $\rm |Q_{33}|$=280\,$^b$ & 1400\,$^a$ &-- & -- & \cite{Yuan2018}\\
    \makecell{0.9PMN-0.1PT\\ceramics} & 2.5\,$^a$ & 500\,$^a$ & 1420\,$^b$ & 0.055\,$^b$ & 18200 \,$^a$& 1040 \,$^a$ & 6.7& \cite{damjanovic1992,swartz1984}\\\hline\hline
    \textbf{Piezoelectrics} & 
    \textbf{\makecell{Maximum field\\(MV/m)}} & 
    \textbf{\makecell{Frequency\\(Hz)}} & 
    \textbf{\makecell{$\rm|M|$\\$(10^{-18}\rm\,m^2/V^2)$}} & 
    \textbf{\makecell{$\rm|Q|$\\$(\rm m^4/C^2)$}} & 
    \textbf{$\varepsilon_r$} & 
    \textbf{\makecell{$\rm |d^{eff}|$ $^\#$\\$(\rm pm/V)$}} &
    \textbf{\makecell{$\rm d^{eff}\cdot g^{eff}$\\$(10^{-12}\rm\,m^2/N)$}} &
    \textbf{Ref}\\\hline
    \makecell{Soft PZT-5H\\ceramics} & 2.4\,$^a$ & 0.2\,$^a$ & -- & -- & 3400 \,$^a$& 750 \,$^a$ & 18.7 & \cite{park1997,jaffe1965}\\
    \makecell{Hard PZT-8\\ceramics} & 4.2\,$^a$ & 0.2\,$^a$ & -- & -- & 1000 \,$^a$& 250 \,$^a$ & 7.1& \cite{park1997,jaffe1965}\\
    \end{tabular}
    
    \begin{tablenotes}
        \footnotesize
        \item[1] $\rm ^a$ is directly mentioned in the reference or has been extracted from the charts, figures or text.
        \item[2] $\rm ^b$ is calculated based on Eqs.\eqref{eq:nonlinear}, \eqref{eq:EPC}, and/or \eqref{eq:MQmq2}
        \item[3] \# $\rm d^{eff}$ is effective piezoelectric coefficients derived from Eq.\ref{eq:EPC} assuming the maximum $dc$ field can be applied or actual piezoelectric coefficient ($\rm d_{33}$) in PZT ceramics.
      \end{tablenotes}
\end{ruledtabular}
\end{table*}

However, these materials were far from being the first ones to be reported as ``giant'' electrostrictors based on their $Q$ values. 
We have classified the reports mentioned above and earlier ones in three general categories: ``giant'' electrostrictors, ``classical'' electrostrictors, and ``not so giant'' electrostrictors: 
\begin{itemize}
    \item ``Giant'' electrostrictors exhibit coefficients more than ten times larger than expected from Eqs.\eqref{eq:trend} or \eqref{eq:Mtrend}, including Gd-doped ceria, (Y, Nb)-stabilized bismuth oxide, undoped LAMOX and lead halide perovskites.
    \item ``Normal'' electrostrictors can exhibit $Q_h$ and $M$ values spanning several orders of magnitude (from $\sim$10$^{-3}$ to $\sim$10$^3$ $\rm m^4/C^2$ for $Q_h$ and from $\sim$10$^{-21}$ to $\sim$10$^{-12}$ $\rm m^2/V^2$ for $|M|$) but correspond within one order of magnitude to the values predicted by Eqs.\eqref{eq:trend} or \eqref{eq:Mtrend}, such as polymers (PVDF etc.), polymer nanocomposites (rGO/PMDS nanocompsites etc.) and classical electrostrictors (PMN-PT etc.).
    \item Following our definitions, there are ``not so giant'' materials that have nevertheless been presented as such. For example: BZT-0.5BCT [\ch{Ba(Zr,Ti)O3}]$_{0.5}$-[\ch{(Ba,Ca)TiO3}]$_{0.5}$ ($Q_{33}$=$0.04\rm\,m^4/C^2$)\cite{jin2016}, (1-x)NN-xBT (1-x)\ch{NaNbO3}-x\ch{BaTiO3} ($Q_{33}$=$0.046\rm\,m^4/C^2$)\cite{zuo2016}, BST \ch{Ba(Sr_x,Ti_{1-x})O3} ($Q_{33}$=$0.0409\rm\,m^4/C^2$)\cite{jin2019high},  NN-BNT \ch{NaNbO3-(Bi,Na)TiO3} ($Q_{33}$=$0.03\rm\,m^4/C^2$)\cite{qi2020}, 
    or CIPS \ch{CuInP2S6} ($Q_{33}$=$-3.2\rm\,m^4/C^2$)\cite{Neumayer2019} for which the $Q$ coefficients are within one order of magnitude of the expected value (still from Eq.\ref{eq:trend}). 
\end{itemize}

A more detailed description of ``giant'' electrostrictors matching our definitions will be presented below.

\subsection{High temperature anionic conductors}
\subsubsection{\ch{Ce_{1-x}Gd_xO_{2-x/2}}}
Gadolinium-doped ceria (CGO) has been applied in the field of solid oxide fuel cells, memristors, sensors, and electrostrictors\cite{esposito2008,sun2012,schweiger2014}. 
Utilizing acceptor dopants like Gd with a 3+ valence state, the fluorite structure of ceria is stabilized with high ionic mobility. 
Gd-doped ceria exhibits giant electrostriction both in thin films and ceramics\cite{korobko2013,Ushakov2017}. 
These electromechanical properties are proposed to originate from the local distortion around oxygen vacancies and its change under applied field\cite{Hadad2016}.
A model of \ch{CeO2} crystalline structure with oxygen vacancies has been proposed to maximize the strain response based on the crystalline orientation\cite{santucci2020}.

The effect of doping, doping percentage, grain boundary, grain size, crystalline orientation, electrodes, measuring methods, and excitation frequency on electrostrictive coefficients have been evaluated; several examples are therefore presented hereafter to give an overview of the giant electrostriction of Gd-doped ceria.

A \ch{Ce_{0.8}Gd_{0.2}O_{1.9}} thin film sandwiched between Cr electrodes and deposited on a glass cantilever was the first to be reported as showing giant electrostriction\cite{Korobko2012}.
A combination of constant and alternating voltages with two harmonic waves at 10\,mHz and 20\,mHz were applied on 400\,nm-700\,nm thin films to ascertain the electrostrictive nature of the response rather than a random vibration.
The curvature of the cantilever was detected through the displacement of laser beam received by a CCD camera. 
From these measurements, we have calculated other coefficients for \ch{Ce_{0.8}Gd_{0.2}O_{1.9}} thin film. They are reported in Table \ref{table:GE}. 
The influence of the oxygen concentration density was then studied in Ref.\onlinecite{Korobko2015}, as different amounts of Gd in the thin film change the oxygen vacancies concentration. 
However, a higher concentration of oxygen does not necessarily result in a larger electrostriction\cite{Korobko2015}. No correlation has been found between the oxygen vacancy concentration and the electrostrictive properties.

Bulk ceramics of Gd-doped ceria coated with gold-silver mixture as electrodes were studied in Ref.\onlinecite{Yavo2018}. 
With the same concentration of oxygen vacancies as in the films, different grain sizes or grain boundaries caused by different synthesis methods lead to different electromechanical performances. 
From Ref.\onlinecite{Kabir2019} 
it appears that the grain size, grain boundary resistivity, and electrostrictive performance are not unambiguously correlated.
Moreover, lower frequency ($<$10\,Hz) excitations may not necessarily lead to higher deformations in ceramics\cite{Yavo2018}. 
A representative example of 10\%mol-doped ceria ceramic measured at low frequency is reported in Table \ref{table:GE}.

\subsubsection{\ch{Bi_7Nb_{2-x}Y_xO_{15.5-x}}}

Cubic bismuth oxide (\ce{Bi2O3}) in the $\delta$ phase adopts a defective fluorite structure with 25\% of the oxygen sites left unoccupied\cite{abrahams2006,wang2006,kaymieva2013}. It is one of the best solid-state ionic conductors with a conductivity around 0.3\,S/cm at 800$^{\circ}$C\cite{Krok2008}.

\ch{Bi_{2}O_{3}} is stable at room temperature in the $\alpha$ (monoclinic) phase, whereas $\delta$-$\rm Bi_{2}O_{3}$ only exists in the range of 729$^{\circ}$C up to melting point 830$^{\circ}$C. 
To stabilize $\delta$-$\rm Bi_{2}O_{3}$ at room temperature, the $\rm Bi_{2}O_{3}$-$\rm Nb_{2}O_{5}$-$\rm Y_{2}O_{3}$ ternary oxide system was selected. 
Isovalent substitution of $\rm Bi^{3+}$ by $\rm Y^{3+}$ produces the disordered fluorite phase with excellent conductivity. 
By contrast, substitution of $\rm Bi^{3+}$ by $\rm Nb^{5+}$ exhibits ordered fluorite phases with much lower ionic conductivity\cite{liu2009}.

Similar to Gd-doped ceria, (Y, Nb)-stabilized $\rm\delta$-$\rm Bi_2O_3$ exhibits giant electrostriction at room temperature, a behavior proposed to be related to the defective fluorite structure and oxygen vacancies\cite{Yavo2016}. As mentioned earlier, such hypothesis was then refuted by the discovery of giant electrostriction in \ch{La2Mo2O9} and related materials that do not adopt the fluorite structure.
With the largest concentration of oxygen vacancies in the investigated $\rm Bi_{2}O_{3}$-$\rm Nb_{2}O_{5}$-$\rm Y_{2}O_{3}$ ternary oxide system, $\rm Bi_7Nb_{0.4}Y_{1.6}O_{13.9}$ was reported\cite{Yavo2016} to have the largest $Q_{33}=-273\rm\,m^4/C^2$ and $M_{33}=-1.2\times10^{-17}\rm\,m^2/V^2$ (see Table \ref{table:GE}). Such values qualify (Y, Nb)-stabilized $\rm\delta$-$\rm Bi_2O_3$ as a ``giant'' electrostrictor as they are thousands of times larger than what would be expected from the empirical scaling law (Eq.\ref{eq:trend}). 


\subsubsection{LAMOX}

Lacorre's team demonstrated \ch{La2Mo2O9} and derivatives (LAMOX) as a distinct structural group of fast oxide-ion conductors\cite{lacorre2000,sellemi2015}, which have been considered as potential core materials for solid oxide fuel cells and oxygen pumping devices.
\ch{La2Mo2O9} (with a conductivity of around 0.06$S/cm$ at 800$^{\circ}$C) can achieve ionic conductivity comparable to commercial 10\% of yttria-stabilized zirconia (YSZ) ($\sim$ 0.1\,S/cm at 1000$^{\circ}$C\cite{etsell1970}) but 200$^{\circ}$C lower. 

This lanthanum molybdate shows an abrupt phase transition from a monoclinic phase ($\alpha$ phase) to a cubic phase ($\beta$ phase) at 580$^{\circ}$C\cite{malavasi2007} and its ionic conductivity increases by almost two orders of magnitudes. 
The slightly distorted monoclinic crystal cell exhibits a large (x2x3x4) superstructure relative to the high temperature cell. 
Non-substituted \ch{La2Mo2O9} ($m$-LAMOX in the reference \onlinecite{Li2018}) contains intrinsic oxygen vacancies without relying on aliovalent doping and Li's group\cite{Li2018} has observed giant electrostriction in this material at 10\,Hz and ambient temperature on ceramics.

Several substitutions can be implemented on both the La and Mo sites of \ch{La2Mo2O9} in order to stabilize the $\beta$ (cubic) form at the operating temperature of fuel cells\cite{georges2003}. 
The $\beta$ phase of tungsten-doped \ch{La2Mo2O9} (\ch{La2MoWO9}, $c$-LAMOX in Ref.\onlinecite{Li2018}) has been investigated at ambient temperature and also exhibits ``giant'' electrostriction.
Dopants such as tungsten are believed to change the activation energy barriers in O-vacancy hopping mechanisms\cite{Li2018}.
Accordingly, the behavior of giant electrostriction would be decided by the nature and concentration of dopants.

A way to consider the high-temperature form of the $\rm\beta$-\ch{La2Mo2O9} is through its structural relationship with $\rm\beta$-\ch{SnWO4}\cite{jeitschko1972}.
In Ref.\onlinecite{Li2018} polarization-electric field ($P$-$E$) and strain-electric field ($x$-$E$) curves have been measured. 
Electrostrictive coefficients are deduced by quadratic fits of the polarization vs electric field and strain vs polarization/E-field curves respectively. 
The magnitude of the electrostrictive coefficients in $m$- and $c$-LAMOX are $M_{33}\sim$10$^{-18}$\,m$^2$/V$^2$ and $Q_{33}\sim$10\,m$^4$/C$^2$ respectively, as shown in Table \ref{table:GE}. With less symmetry distortion, $c$-LAMOX is slightly less electrostrictive than $m$-LAMOX despite its higher permittivity. 
The occurrence of giant electrostriction in LAMOX demonstrates that this phenomenon can exist in non-fluorite structures.

\subsubsection{Conclusion for high-temperature anionic conductors}

As a conclusion, giant electrostriction has been measured in ceramics and thin films of ionic conductors. 
Contrary to what was initially believed, the phenomenon is not confined to the fluorite structure. 
All of them, though, have intrinsic oxygen vacancies and the current hypothesis is that the global strain developed under electric field is related to the field-response of the large local strain induced by the presence of oxygen vacancies. 
This would, however, indicate that a larger concentration of oxygen vacancies lead to a larger electrostrictive response (as long as they do not interact), which is not what reports on Gd-doped ceria\cite{Yavo2018} indicate.
The relationship between the oxygen vacancy concentration and the overall deformation remains to be unravelled.
Another common feature of the electrostrictive response of these materials is that it occurs at lower frequency than ``classical'' electrostriction (around 10\,Hz), with a sharp decrease when the frequency of the applied field increases. 
Such low threshold frequency is consistent with electro-active species heavier than electrons, such as oxygen vacancies.
In addition, Gd-doped ceria\cite{kraynis2017} $\delta$-\ch{Bi2O3}\cite{Yavo2016} as well as LAMOX\cite{fang2006} exhibit anelastic behavior. The nature of the interaction between elastic and electric dipoles may therefore be the key to understand giant electrostriction in these materials. 

\subsection{Lead halide perovskites}
The initial interest in lead halide perovskites laid in photovoltaics, photodetectors, radiation detectors, and light-emitting diodes (LED) due to their high power conversion efficiency (\textgreater20\%), tunable bandgap, large mobilities of electrons and hole, and low fabrication cost\cite{yang2017,xiao2017}. They have since  also attracted attention for their electromechanical properties\cite{chen2018}.

In Ref.\onlinecite{chen2018}, three different samples are reported to exhibit electrostrictive responses: inorganic lead perovskite \ch{CsPbI_3} single crystal (500\,nm-thick) exhibits a +0.2\% strain under 5\,MV/m $ac$ field at 10\,Hz ($M=-8\times10^{-17}\rm\,m^2/V^2$); organic–inorganic hybrid perovskite (OIHPs) \ch{MAPbBr3} single crystal (95\,$\rm\mu$m-thick) exhibits a -0.36\% strain under 2\,MV/m at 10\,Hz ($M=-9\times10^{-16}\rm\,m^2/V^2$); organic–inorganic hybrid perovskites \ch{MAPbI3} single crystal (40\,$\rm\mu$m-thick) exhibits under 3.7\,MV/m at 100\,Hz a maximum negative strain reaching -1\% with $M=-7.3\times10^{-16}\rm\,m^2/V^2$. 
Their electrostrictive performances are stable either under 1200 $ac$ loops actuation or 60\,s of square waves. 
Unlike the previous oxides, lead halide perovskites do not contain oxygen vacancies. 
It is speculated that lattice deformation is caused by the formation of additional defects.

Several possible alternative mechanisms to electrostriction were discarded in these materials, such as Maxwell strain, organic dipoles, local polar fluctuations, or ferroelectricity.
The depolarization field needed to yield a Maxwell strain caused by ion accumulation would have to be 527\,MV/m, which is more than two orders magnitudes larger than the applied field (3.7\,MV/m).
The existence 
organic dipoles 
is excluded due to the electrostrictive (rather than piezoelectric) response of \ch{CsPbI_3}.
Local polar fluctuations occur at the picosecond or a few hundred femtosencond timescale, which is much faster than the electromechanical response of \ch{MAPbI3}. 
The electromechanical performance of \ch{MAPbBr3} rules out the contribution from ferroelectricity since \ch{MAPbBr3} single crystal has a centrosymmetric structure and thus cannot be ferroelectric.

Lead halide perovskite may therefore constitute a disctinct group of ``giant'' electrostrictors as they do not contain (a large number of) oxygen vacancies. Their electrostrictive response nevertheless occurs at relatively low frequency, a characteristics shared with the one of ionic conductors that are ``giant'' electrostrictors. This suggests that the electro-active species differ from the ones of classical electrostrictors. Similarly to ionic conductor-based electrostrictors, anelastic relaxation was reported at 0.48, 2.5 and 6.1 \,kHz\cite{cordero2018} in \ch{MAPbI3}.
The presence of lead, though, represents a major drawback for their potential applications as electromechanical devices.


In addition to ceramics and thin films of inorganic or organic-inorganic materials, giant electrostrictive responses have also been reported in polymers and polymer-based composites. Such electrostrictors are reviewed below. 

\subsection{Polymer nanocomposites}

Flexible electrostrictive materials have numerous applications, \textit{e.g.} in wearable sensors and MEMS\cite{zhao2019flexible,nesser2019harvesting,rajitha2018optically}.
Inspired by the introduction of dielectric mismatched nanodomains in a dielectric elastomer\cite{kim2011electric} and electron-irradiated poly(vinylidene fluoride-trifluoroethylene) [P(VDF-TrFE)] copolymer\cite{Zhang1998}, a large electrostriction has been observed\cite{Yuan2018} in a composite of reduced graphene oxide (rGO) embedded in a polydimethylsiloxane (PDMS) matrix. These nanocomposites (rGO/PDMS) combine a ``giant'' polarization electrostriction coefficient $Q$, the large elastic compliance $s$ from polymers, and a large dielectric contribution due to the surface defects from rGO, which bring about ultra-large field electrostriction coefficient $M$ (see Table \ref{table:GE}) despite not qualifying them as ``giant'' $M$ electrostrictors (see Fig.\ref{fig:Mse}).

Locally aligned structure of PDMS contains 4.7wt\% of rGO flakes with large anisotropy. 
The nanocomposites have large permittivity (around 1500 at 0.1\,Hz), which is attributed to liquid crystals with long-range orientational order\cite{Yuan2018}. 
The ordered structure is expected to promote the formation of extended microcapacitors. 

The permittivity of pure PDMS slightly decreases as a function of strain\cite{Yuan2018}. The field electrostrictive coefficient ($M$) and relative permittivity of this  polymer are around $-1.5\times10^{-18} \rm\, m^2/V^2$ and 2.8 at 100\,Hz, respectively.
In contrast, rGO/PDMS composites exhibit a steep decrease in the permittivity-strain curve. 
The rGO doping therefore improves the electrostrictive responses based on Eqs. \eqref{eq:MQmq} and \eqref{eq:MQmq2}. 
The reported field electrostrictive coefficient is around $-1.2\times10^{-14} \rm\, m^2/V^2$ at 0.1\,Hz from the permittivity-strain curve. 
The relative permittivity reaches 1500 without applied strain. 
The dielectric spectroscopy used in the experiment enables to avoid the contribution of otherwise non-negligible Maxwell stress effects. 
Indeed, for this nanocomposite, the Maxwell strain coefficient ($s\epsilon_0\epsilon_r/2$) is $-3.1\times10^{-15} \rm\, m^2/V^2$, only one order of magnitude lower than the electrostrictive field coefficient  $M\sim-1.2\times10^{-14}\rm\,m^2/V^2$.
In contrast, \textit{e.g.} in bismuth oxide, the Maxwell strain coefficient is $-2.5\times10^{-21}\rm\,m^2/V^2$, which is negligible compared to the electrostriction coefficient $M\sim-1.2\times10^{-17}\rm\,m^2/V^2$. 
Therefore, applying high voltages to composites will always lead to a strain combining the electrostrictive and electrostatic responses.

Another type of composites investigated for their electrostrictive properties is the carbon nanotubes CNT-PDMS composites. 
CNT concentration varies from 0.05\,wt\% to 2\,wt\%, covering the percolation threshold around 1\,wt\%. 
At 100\,Hz, the permittivity of CNT composites rises with increasing concentration of CNT. 
$M_{33}$ demonstrates the same increasing trend except that the  deformation of 2\,wt\% CNT composites switches sign (from negative to positive). The maximum negative $M_{33}$ is obtained at 1\,wt\%, \textit{i.e.} at the percolation threshold.
$Q_{33}$ is in the range of $-6\times10^3\rm\,m^4/C^2$ to $-1.2\times10^5\rm\,m^4/C^2$. In contrast to $M$, the maximum negative $Q_{33}$ is attained at 0.5\,wt\%, \emph{i.e.} below the percolation. 
 
As a conclusion, ``giant'' electrostrictors based on polymers and composites exhibit maximum strains that are much larger than inorganic ones (several percents versus a few 0.1\%), field electrostriction coefficient ($M$) are several orders of magnitude larger than their inorganic counterparts but do not qualify them as ``giant'' electrostrictors as their performance are comparable with the expectation from Eq.\eqref{eq:Mtrend}.  
 
\section{Interests of giant electrostrictors}

The discovery of functional materials exhibiting responses much larger than anticipated is always of interest. 
From a fundamental point of view, understanding the origin of this response would enable to engineer dedicated structures to further enhance this property. 
From an applied point of view, such a large electromechanical response offers the perspective of an alternative to piezoelectric materials that have the drawback of being lead-based in the majority of applications. 
However, before these materials revolutionize the market of electro-mechanical devices, several major advances must be made. We review the most pressing ones below.

\subsection{Fundamental interest}

According the prevailing hypothesis, ``giant'' electrostriction is related to the field-dependence of the local strain induced by oxygen vacancies in inorganic materials. 
The ``active ingredient'' would then differ from the electrons of classical electrostriction. 
The exact origin of electrostriction in inorganic materials remains to be determined before structures can be tailored to harness its full potential. 

The relation between oxygen vacancy concentration and electrostrictive properties also needs to be clarified. 
The observed increase of the response amplitude with oxygen vacancies concentration in (Nb,Y)-doped \ch{Bi2O3}\cite{Yavo2016} 
is in contradiction with the results on Gd-doped ceria\cite{Korobko2012}.
Such increase may only occur as long as the vacancies do not interact with each other. 
The exact role of oxygen vacancies in ``giant'' electrostriction remains to be understood.

The fact that the few materials discovered so far exhibit such a response, already vastly surpassing their classical counterparts, suggests that more improvement can still be made. 
Such an improvement lead the piezoelectric perovskites from \ch{BaTiO3} ($d_{33}=85.6\rm\,pm/V$ single crystals\cite{berlincourt1958}) to PZT ($d_{33}=750\rm\,pm/V$\cite{jaffe1965}) with a tenfold improvement.  
If ``giant'' electrostriction is indeed related to local strain relaxation in the vicinity of an oxygen vacancy induced by an aliovalent dopant, then the height of the energy barrier preventing the oxygen vacancy to jump from one oxygen site to another should be as high as possible, thereby inhibiting ionic conduction. 
This suggests that even larger electrostrictive properties may be found in oxygen-deficient materials with low ionic conductivity. 

As a consequence, the study of dopants having a strong affinity for the oxygen vacancies they induce represents a potentially promising path to increase the electromechanical response.

It also remains to be clarified whether the other family of ``giant'' electrostrictors (lead halide perovskites and polymer nanocomposites. We concluded polymer nanocomposites were not true giant ES) present a similar mechanism (local, field-dependent strain) or not. In order to describe and predict such behavior, an appropriate effective medium theory could be of great help. 

From this understanding of the material properties that give rise to ``giant'' electrostrictive responses, the possibility arises to tune structures to exacerbate this response further. Besides, literature survey or databases exploration could point out to materials already available that have not been considered so far for their electrostrictive properties. 

Unravelling the fundamental mechanism(s) giving rise to what we define as ``giant'' electrostriction may enable to differentiate materials following Eqs.\eqref{eq:trend} and/or \eqref{eq:Mtrend} from the others. If materials can be grouped by their electrostrically active species, it may then be preferable to abandon superlatives altogether in favor of epithets referring to the underlying mechanism.

\subsection{Interest for applications}

Electrostriction is of interest for both actuation and sensing applications\cite{damjanovic1992}. 
The strain response usually exhibits low hysteresis, is weakly temperature dependent and exists in all dielectrics. 
In addition, electrostrictive materials do not require poling and therefore exhibit superior fatigue behavior. 
As such, it offers a potential alternative to piezoelectrics.
The modest performances of classical inorganic electrostrictors have so far restricted electrostrictors to niche markets. 
The discovery of ``giant'' electrostrictors may trigger a renewed interest in the application potential of these materials.
Apart from the amplitude of the response, a major difference between classical inorganic electrostrictors and their ``giant'' counterparts lies in the cutoff frequency of ``giant'' electrostriction. 
Giant electrostrictors exhibit their highest responses in the 10$^{-2}$\,Hz to 10$^{3}$\,Hz range. 
Their application to high frequency devices is therefore not possible at the present time. 

All the ``giant'' electrostrictors discovered so far contract under electric field, whereas classical relaxors or ferroelectrics usually expand.
This means that Maxwell strain will actually be beneficial to these materials, contributing to larger negative strain instead of decreasing positive strains. ``Giant'' electrostrictors may therefore have larger deformation under the field thanks to this electrostatic attraction.

Transduction coefficient ($\rm d_{33}\cdot g_{33}$) of the piezoelectric energy harvesting materials is another potential field of interest for ``giant'' electrostrictors.
It is difficult to obtain simultaneously high piezoelectric coefficient $d_{33}$ and low permittivity $\varepsilon$ using doping or composites\cite{yu2019}.
``Giant'' electrostrictors usually present low permittivities and high effective piezoelectric coefficients, as shown in the Table \ref{table:GE}. This may provide an avenue to improve the performance of energy harvesting devices, especially those harnessing ambient (\emph{i.e.} low frequency) vibrations.

There are, however, several hurdles to overcome. 
The first one is not related to the materials themselves but to the electronic environment of these sensors and actuators. 
Indeed, such systems have been carefully optimised for linear electromechanical responses. 
The quadratic nature of the electrostrictive response may therefore pose a challenge, unless electrostrictors are used as effective piezoelectrics. 
This would impose to control their dielectric losses as bias voltages would have to be imposed. 
Measuring and enhancing their breakdown voltage is another crucial aspect of the applicability of these materials. 
This is not restricted to ``giant'' electrostrictors, though, as the effective piezoelectric coefficient of PMN-0.1PT (a ``classical'' electrostrictor) can reach 1500\,pm/V under a bias field of 3.7\,kV/cm. This value is 2-3 times larger than the one of PZT ceramics ($\sim$ 400\,pm/V) under a field of 3 kV/cm\cite{damjanovic1992}.
Finding electrostrictors with $M$ coefficients larger that 10$^{-15}$m$^2$/V$^2$ (the upper current limit of relaxor perovskites that are ``normal'' electrostrictors) would bring effective piezoelectric coefficients thousands of times larger than PZT ceramics.

The second hurdle about giant electrostrictors is related to the stability of the response. 
Thorough investigations of the fatigue and degradation behavior of inorganic giant electrostrictors are still missing. 
Such information is critical for the viability of the devices based on these materials. 

A third hurdle is related to the current spread of reported values for the electrostriction coefficients, even for the same material. For example in Ref.\onlinecite{Yavo2018}, electrostrictions are presented by samples obtained with different sintering procedures. The low frequency longitudinal electrostrictive strain coefficient $M_{33}^0$ of these samples vary from $-2.38\times10^{-17} \rm m^2/V^2$ to $-21.7\times10^{-17} \rm m^2/V^2$. 
Such variations suggest that the microstructure of the samples play a role in the response (as is also the case for piezoelectrics). A better understanding of the role of the microstructure on the final properties would help designing synthesis methods ensuring more consistent performances. 

\section{Conclusion}

Electrostriction is an electromechanical phenomenon that exists in all dielectrics. 
In the direct effect, the induced strain or stress is proportional to the square of the polarization or the electric field (in linear dielectrics). 
Reverse effects exists as well, provided the sample is pre-strained or polarised. 
This enables both actuation and sensing. 
A pseudo-piezoelectric effect (linear response at the same frequency as the excitation) can be induced under a bias field much larger than the excitation field. 

The applications of electrostriction have so far been restricted to niche markets due to the superior performances of piezoelectrics. 
The recent discovery of ``giant'' electrostrictors has the potential to challenge this state. 

We propose to define as ``giant'', electrostrictors that exhibit electrostriction coefficients at least ten times larger than the value expected either from the empirical relation put forward by Newnham\cite{R.E.Newnham} for the polarization ($Q$) coefficient or the one we posit for the field ($M$) coefficient. 
The latter is of most interest for applications as it relates the induced strain to the electric field. 
We also demonstrate that a ``giant'' $Q$ coefficient does not necessarily ensure a ``giant'' $M$ coefficient. 

From these definitions, we review the existing literature on electrostrictors and classify them as indeed ``giant'', ``normal'', or ``not so giant'' (\emph{i.e.} ``normal'' despite having been reported as ``giant''). We also underline the fact that the electrostriction coefficients of ``giant'' electrostrictors are \textit{anomalously} large (compared to the expected values) rather than very large in \textit{absolute} values (compared to all other electrostrictors, including ``normal'' ones).

An extensive review of the ``giant'' electrostrictors enables to underline their common characteristics: the existence of point defects (\textit{e.g.} oxygen vacancies) and the restriction of the ``giant'' response to low frequencies. Such a response is not limited to a particular structure nor to the organic, inorganic, or mixed nature of the material. It even extends to composites. 

Much work remains to be done to understand the fundamental mechanism or mechanisms at play in ``giant'' electrostriction. For example, the exact role of point defects, the relationship between their density and the amplitude of the electrostrictive response, the role of elastic dipoles and microstructure \textit{etc.} need to be clarified to understand how to engineer materials able to exhibit even larger responses. 

Such an understanding is a pre-requisite to the replacement of (too-often) lead-based piezoelectric materials by environmentally friendlier electrostrictors with superior, reliable, and optimised properties.

\begin{acknowledgments}
This work was supported financially by the ANR-19-ASTR-0024-01 and ANR-20-CE08-0012-1 grants, as well as the Chinese Scholarship Council. We thank P. Lacorre for the discussions. 
\end{acknowledgments}

\bibliography{ref}

\end{document}